\documentclass[aps,prd,superscriptaddress,nofootinbib,amsmath,amsfonts,preprintnumbers,groupedaddress,showpacs,10pt,english]{revtex4-1}
\usepackage{amsmath}
\usepackage{amssymb}
\usepackage{babel}
\usepackage{wrapfig}
\usepackage{cancel}
\newcommand{\nn}{\nonumber \\}

\makeatletter

\usepackage{array,multirow,graphicx}
\usepackage{dcolumn}
\usepackage{newlfont}
\usepackage{bm}
\usepackage[colorlinks,citecolor=blue,urlcolor=blue,linkcolor=blue]{hyperref}
\usepackage[figtopcap]{subfigure}
\usepackage{color}

\begin{document}

\date{\today}

\title{Non-trivial black hole solutions in $\mathit{f(R)}$ gravitational theory}

\author{G.G.L. Nashed}
\email{nashed@bue.edu.eg}
\affiliation {Centre for Theoretical Physics, The British University, P.O. Box
43, El Sherouk City, Cairo 11837, Egypt}

\author{S. Nojiri}
\email{nojiri@gravity.phys.nagoya-u.ac.jp}
\affiliation{Kobayashi-Maskawa Institute for the Origin of Particles and the Universe, Nagoya University, Nagoya 464-8602, Japan}

\begin{abstract}
 Recent observation shows that general relativity (GR) is not valid in the strong regime.  $\mathit{f(R)}$ gravity where  $\mathit{R}$ is the Ricci scalar, is regarded to be one of good candidates able to cure the anomalies appeared in the conventional general relativity. In this realm, we apply the equation of motions of $\mathit{f(R)}$ gravity to a spherically symmetric spacetime with two unknown functions and derive original black hole (BH) solutions without any constrains on the Ricci scalar  as well as on the form of $\mathit{f(R)}$ gravity. Those solutions depend on a convolution function and are deviating from the Schwarzschild solution of the Einstein GR. These solutions are characterized by the gravitational mass of the system and the convolution function that in the asymptotic form gives extra terms that are responsible to make such BHs different from GR. Also, we show that these extra terms make the singularities of the invariants much weaker than those of the GR BH. We analyze such BHs using the trend of thermodynamics and show their consistency with the well known quantities in thermodynamics like  the Hawking radiation, entropy and quasi-local energy. We also show that our BH solutions satisfy the first law of thermodynamics.  Moreover, we study the stability analysis using the odd-type mode and shows that all the derived BHs are stable and have radial speed equal to one. Finally, using the geodesic deviations we derive the stability conditions of these BHs.
\end{abstract}

\pacs{04.50.Kd, 04.25.Nx, 04.40.Nr}
\keywords{$\mathbf{F(R)}$ gravitational theory, analytic spherically symmetric BHs, thermodynamics, stability, geodesic deviation.}

\maketitle
\section{\bf Introduction}
 More than one decade ago the Newton gravity failed to investigate some issues like the advances of Mercury besides  the Mickelson Merry experiment \cite{PhysRevLett.103.090401}. In 1915, Einstein constructed his famous theory, the general theory of relativity (GR), that was able to resolve the issue of Mercury \cite{Wheeler:1990nd}. After that most researchers trust  GR as a successful theory of the gravitational field. However, in recent time this theory, GR, failed to be consistent with observation and was not able to describe the dark energy and dark matter that are confirmed by observations \cite{Perlmutter:1998np,Riess:1998cb,Riess:2004nr,Hirata:1987hu,Dodelson:1993je,Cole:1994ab}. Also GR gives a violation of the Chandrasekhar mass-limit for white dwarfs of the super-Chandrasekhar as well as white dwarfs of the sub-Chandrasekhar limiting mass \cite{Howell:2006vn,Scalzo:2010xd,Filippenko:1992wda,10.1093/mnras/284.1.151,Turatto:1998eq,2001PASP..113..308M,Garnavich:2001vx,Taubenberger:2007dt}.

The Einstein theory of GR has many forms of modifications, like Also modifications of GR action can be achieved to derive different
kinds of modification theories of gravity like $f(T)$ gravity
\cite{Cai:2015emx,Awad:2017yod,Awad:2017ign,Zubair:2015cpa}, where $T$ is the torsion scalar in teleparallelism.
$f(R)$ gravity \cite{DeFelice2010,Nojiri:2010wj,Capozziello:2011et,Nojiri:2017ncd,Capozziello:2010zz} with $R$ the scalar curvature; $f(G)$ gravity with $G$ the Gauss-Bonnet invariant \cite{Cognola:2006eg}; $f(R,{\cal T})$ gravity, where ${\cal T}$ the trace of the energy-momentum tensor of matter \cite{Harko:2011kv,2016Ap&SS.361....8Z}; $f(T,{\cal T})$ gravity, where $T$ is the torsion scalar in teleparallelism and ${\cal T}$ the trace of the energy-momentum tensor of matter \cite{Saleem:2020acy} etc. All of these modified theories  have received much attention to investigating the shortcomings  of GR like the accelerated expansion of our universe; investigation of flat rotation curves of galaxies;  wormhole behavior and another ambiguous phenomenon near BHs \cite{DeFelice:2010aj,CAPOZZIELLO2011167,Nojiri:2006ri,universe1020123,Bamba:2012cp}. The first time to use a quadratic form of the Ricci scalar was given by Starobinsky  \cite{1979ZhPmR..30..719S}. It was shown that the higher-order curvature of $\mathit{f(R)}$ gravity  can solve the issue of the massive neutron stars \cite{Astashenok_2013,PhysRevD.89.103509,Astashenok_2015,Astashenok_2017,Astashenok:2016epm}. It is well-known that  $\mathit{f(R)}$ gravity  is  consists of an arbitrary function whose first order is the Ricci scalar. The equation of motions of $\mathit{f(R)}$ gravity have  higher degrees and supply substantial classes of  solutions that are different from GR. In the frame of $\mathit{f(R)}$ gravity the dynamical behavior of the matter field and dark energy has been studied \cite{6847167,Capozziello_2013,10.1093/pasj/psx099,PhysRevD.74.046004}. From the viewpoint of cosmology, many researchers have been carried out their studies from different directions \cite{Shah:2019mxn,Nojiri:2019dqc,Odintsov:2019ofr,PhysRevD.99.064049,Nascimento:2018sir,Miranda:2018jhu,Astashenok:2018bol,PhysRevD.99.063506,Elizalde:2018now,PhysRevD.99.064025,PhysRevD.99.104046,
Bombacigno:2018tyw,Capozziello:2018ddp,Samanta:2019tjb}.
A  spherically symmetric vacuum BH solution in  $\mathit{f(R)}$  gravity has been  derived in \cite{PhysRevD.74.064022,2018EPJP..133...18N,2018IJMPD..2750074N,Nashed:2018piz}. Using Noether symmetry Capozziello et al. have derived spherically symmetric solutions  in the frame of  $\mathit{f(R)}$  gravity \cite{Capozziello_2007,2012GReGr..44.1881C}. Using the same techniques, Noether symmetry, axially symmetric vacuum BH solutions are derived \cite{Capozziello_2010}. Non-trivial spherically symmetric BH solutions for a specific class of  $\mathit{f(R)}$ gravity are derived \cite{Elizalde:2020icc,Nashed:2019yto,Nashed:2019tuk}. Due to the fact of the existence of higher-order curvature terms in $\mathit{f(R)}$ gravity one can discuss potentially the importance of strong gravitational background in local objects. In this frame, many researchers are concentrate  to study spherically symmetric, static BHs \cite{Sultana:2018fkw,Canate:2017bao,Yu:2017uyd,Canate:2015dda,Kehagias:2015ata,PhysRevD.82.104026,delaCruzDombriz:2009et}
and Neutron stars \cite{Feng:2017hje,Resco:2016upv,Capozziello:2015yza,Staykov:2018hhc,Doneva:2016xmf,Yazadjiev:2016pcb,Yazadjiev:2015zia,Yazadjiev:2014cza,
Ganguly:2013taa,Astashenok:2013vza,Orellana:2013gn,Arapoglu:2010rz,Cooney:2009rr} solutions in the quadratic model of $\mathit{f(R)}$ gravity. We note that $\mathit{f(R)}$ gravity
is equivalent to the Brans-Dicke   theories \cite{PhysRev.124.925} with a scalar potential of the gravitational origin \cite{Chiba:2003ir,PhysRevLett.29.137,Chakraborty:2016gpg,Chakraborty:2016ydo}. It is the purpose  of this manuscript  to derive  original spherically symmetric BHs in $\mathit{f(R)}$ gravity without assuming any constrains on Ricci scalar nor on the  form of $\mathit{f(R)}$ gravity and study the relevant physics of those BHs.

 The structure of this study is as follows: In Sec. \ref{S2} we give the fundamentals of  $\mathit{f(R)}$  gravity. In Sec. \ref{S3} we apply the field equations of $\mathit{f(R)}$ gravity to a spherically symmetric line-element having unequal metric potentials. We derive the system of differential equations that have three unknown functions and derive different solutions of this system that is characterized by a convolution function. If this convolution function is vanishing we return to the BH of GR, The Schwarzschild solution. So this convolution function appears as the effect of higher-order curvature that characterizes $f(R)$ gravity. Moreover, we give the asymptote form of this convolution function up to certain order and show the trace of the higher-order curvature. Also, we calculate  the Kretschmann scalar, the Ricci tensor square and the Ricci scalar,  and show the trace of $\mathit{f(R)}$  gravity on such invariants that makes the singularity weaker than those of GR BHs.  In Sec. \ref{S4} we calculate the above mentions thermodynamical quantities to be compared with their previous findings. In Sec. \ref{S616} we use the odd-type method and study the stability of these BHs. Also in Sec. \ref{S616}, we use the geodesic deviation to derive the condition of stability for such BHs derived in  Sec. \ref{S3}. In the final section we give our concluding remarks.

\section{Fundamentals of $\mathit{f(R)}$ gravitational theory}\label{S2}
In this section, we consider a 4-dimensional action of  $\mathit{f(R)}$ gravity where    $\mathit{f(R)}$ is an arbitrary differential function. It is important to stress on the fact that   $\mathit{f(R)}$ gravity is a modification of  GR and coincides with the Einstein GR at lower order, i.e.,   $\mathit{f(R)=R}$. When  $\mathit{f(R)\neq R}$  then we have a theory that is different from GR. The action  of $\mathit{f(R)}$  gravity  can take the form (cf. \cite{Carroll:2003wy,1970MNRAS.150....1B,Nojiri:2003ft,Capozziello:2003gx,Capozziello:2011et,Nojiri:2010wj,Nojiri:2017ncd,Capozziello:2002rd}):
\begin{eqnarray} \label{a2} {\mathop{\mathcal{ I}}}:=\frac{1}{2\kappa} \int d^4x \sqrt{-g}  \mathit{f(R)}\,,\end{eqnarray}
where $\kappa$ is  Newton's gravitational constant and  $g$ is the determinant of the metric.

The use of variations principle  to action (\ref{a2}) gives  the vacuum field equations to become \cite{2005JCAP...02..010C}
\begin{eqnarray} \label{f1}
{\mathop{\mathcal{ I}}}_{\mu \nu}=\mathit{ R}_{\mu \nu} \mathit{f_{R}}-\frac{1}{2}g_{\mu \nu}\mathit{f( R)}+[g_{\mu \nu}\Box -\nabla_\mu \nabla_\nu]\mathit{ f}_{_\mathit{ R}} \equiv0,\end{eqnarray}
such that  $\Box$ is the d'Alembertian operator and $\displaystyle  \mathit{f_{R}}=\frac{\mathit {df}}{\mathit {dR}}$.
The trace of the field equations ~(\ref{f1}), takes the form:
\begin{eqnarray} \label{f3}
{\mathop{\mathcal{ I}}}=3\Box {\mathit f_{R}}+\mathit{ R}{f_{R}}-2\mathit f(R)\equiv0 \,.\end{eqnarray}
From Eq. (\ref{f3}) one can obtain $\mathit f(R)$  in the form:
\begin{eqnarray} \label{f3s}
\mathit f(R)=\frac{1}{2}\big[3\Box {\mathit f_{R}}+\mathit{ R}{f_{R}}\Big]\,.\end{eqnarray}
Using Eq. (\ref{f3s}) in Eq. (\ref{f1}) we get \cite{Kalita:2019xjq}
\begin{eqnarray} \label{f3ss}
{\mathop{\mathcal{ I}}}_{\mu \nu}=\mathit{ R}_{\mu \nu} \mathit{f_{R}}-\frac{1}{4}g_{\mu \nu}\mathit{ R}\mathit{ f}_{_\mathit{ R}}+\frac{1}{4}g_{\mu \nu}\Box\mathit{ f}_{_\mathit{ R}} -\nabla_\mu \nabla_\nu\mathit{ f}_{_\mathit{ R}}=0  \,.\end{eqnarray}
Accordingly, it is well be important to examine Eqs. (\ref{f3}) and (\ref{f3ss}) to a spherically symmetric spacetime having two unknown functions.
\section{Spherically symmetric BH solutions}\label{S3}
   To study the  equation of motions (\ref{f3}) and (\ref{f3ss})  in order to derive a general form of the arbitrary function $\mathit{f(R)}$ without assuming any restrictions on the Ricci scalar we use a spherically symmetric spacetime having two unknown functions of the following form:
\begin{eqnarray} \label{met12}
& &  ds^2=-A(r)dt^2+\frac{dr^2}{B(r)}+r^2(d\theta^2+\sin^2\theta d\phi^2)\,,  \end{eqnarray}
with $A(r)$ and $B(r)$ are  functions depending on  radial coordinate $r$. The Ricci scalar of the metric (\ref{met12})  figured out as:
  \begin{eqnarray} \label{Ricci}
  {\textit R(r)}=\frac{r^2BA'^2-r^2AA'B'-2r^2ABA''-4rA[BA'-AB']+4A^2(1-B)}{2r^2A^2}\,,
  \end{eqnarray}
  where $A\equiv A(r)$,  $B\equiv B(r)$,  $A'=\frac{dA}{dr}$, $A''=\frac{d^2A}{dr^2}$ and $B'=\frac{dB}{dr}$.
  Plugging Eqs.
 (\ref{f3}), (\ref{f3ss}) with Eq. (\ref{met12}) and by  using Eq. (\ref{Ricci}) we get:
 \begin{eqnarray}
&& {\mathop{\mathcal{ I}}}_t{}^t=\frac{1}{8r^2N^2}\Bigg\{r^2[BFN'^2-3NFB'N'-2NBFN''-2N^2FB''-3NBN'F'-2N^2B'F'+2BN^2F'']\nonumber\\
&&-4rNB[FN'-NF']-4N^2F[1-B]\Bigg\}=0\,,\nonumber\\
&&{\mathop{\mathcal{ I}}}_r{}^r=\frac{1}{8r^2N^2}\Bigg\{r^2[BFN'^2-3NFB'N'-2NBFN''-2N^2FB''+NBN'F'-2N^2B'F'-6BN^2F'']\nonumber\\
&&+4rNB[FN'+NF']-4N^2F[1-B]\Bigg\}=0\,,\nonumber\\
&&{\mathop{\mathcal{ I}}}_\theta{}^\theta={\mathop{\mathcal{ I}}}_\phi{}^\phi=\frac{1}{8r^2N^2}\Bigg\{r^2[3NFB'N'+2NBFN''+2N^2FB''-BFN'^2+NBN'F'+2N^2B'F'+2BN^2F'']\nonumber\\
&&-4rN^2BF'+4N^2F[1-B]\Bigg\}=0\,,\nonumber\\
&&{\mathop{\mathcal{ I}}}=\frac{1}{2r^2N^2}\Bigg\{r^2[6N^2B'F'-3NFB'N'-2NBFN''-2N^2FB''+BFN'^2+3NBN'F'+6BN^2F'']\nonumber\\
&&+4rN[3NBF'-FBN'-2FNB']+4N^2F[1-B]-4r^2N^2f(r)\Bigg\}=0\,,
\label{feq}
\end{eqnarray}
where $N(r)=\frac{A(r)}{B(r)}$ and $F\equiv F(r)=\frac{df(R(r))}{dR(r)}$, $F'=\frac{dF(r)}{dr}$, $F''=\frac{d^2F(r)}{dr^2}$, $F'''=\frac{d^3F(r)}{dr^3}$. Since we are dealing with spherical symmetry we take $f(R)=f(r)$. It is of interest to mention here that the above system of differential equations given by (\ref{feq}) is identical with the differential equations given in \cite{Jaime:2010kn}.

Equations (\ref{feq}), except the trace part, can be rewritten in the following form;
\begin{align}
\label{E1n}
0=& r^2\left[BF{N'}^2 - 3NFB'N' - 2NBFN'' - 2N^2FB'' - 3NBN'F' - 2N^2B'F'
+ 2BN^2F'' \right] \nn
& - 4rNB \left[ FN' - NF' \right] - 4N^2F \left[1 - B \right] \, , \\
\label{E2n}
0=&r^2 \left[ BF{N'}^2 - 3NFB'N' - 2NBFN'' - 2N^2FB'' + NBN'F'  - 2N^2B'F'
 - 6BN^2F'' \right] \nn
& +4rNB \left[FN'+ NF' \right] - 4N^2F \left[1 - B\right] \, , \\
\label{E3n}
0=& r^2 \left[ - BF{N'}^2 + 3NFB'N' + 2NBFN'' + 2N^2FB'' + NBN'F' + 2N^2B'F'
+ 2BN^2F'' \right] \nn
& - 4rN^2BF' + 4N^2F \left[1 - B\right] \, .
\end{align}
By using Eqs. (\ref{E1n}) and (\ref{E2n}), ((\ref{E1n}) minus (\ref{E2n})), we obtain
\begin{equation}
\label{E4n}
0 = r^2 \left[ - 4NBN'F' + 8BN^2 F'' \right] - 8r NN'B F \, .
\end{equation}
On the other hand by using Eqs. (\ref{E1n}) and (\ref{E3n}), ((\ref{E1n}) plus (\ref{E3n})),
we obtain
\begin{equation}
\label{E6n}
0 = - 2 r^2 NBN'F' + 4r^2BN^2 F'' - 4r NN' BF \, ,
\end{equation}
which is identical with (\ref{E4n}) and therefore
only two equations in (\ref{E1n}), (\ref{E2n}), and (\ref{E3n}) are independent.
For example Eq.~(\ref{E1n}) is equal to minus Eq.~(\ref{E2n}) minus two times Eq.~(\ref{E3n}).
Then for example we can choose Eq.~(\ref{E1n}) and Eq.~(\ref{E6n}) as independent
equations.
Because we have three unknown functions $B$, $N$ and $F$, we cannot determine one function.

As an example, we assume the Schwarzschild type solution,
\begin{equation}
\label{ES1}
N=1 \, .
\end{equation}
By assuming $B\neq 0$ almost everywhere, which is a physically required, Eq.~(\ref{E4n}) gives
\begin{equation}
\label{ES2n}
F''=0 \, , \quad \mbox{that is,} \quad
F=F_0 + F_1 r \, .
\end{equation}
Eq.~(\ref{E1n}) has the following form
\begin{align}
\label{ES3n}
0=& r^2\left[ - 2 FB'' - 2 B'F' \right] + 4r B F' - 4F \left[1 - B \right] \nn
=& - 2 r^2 \left( F_0 + F_1 r \right) B'' -2 r^2 F_1 B' + 4 \left( F_0 + 2 F_1 r \right) B
 - 4 \left( F_0 + F_1 r \right) \, .
\end{align}
In case $F_1=0$, Eq.~(\ref{ES3n}) reduces to
\begin{equation}
\label{ES4n}
0= - r^2 B'' + 2 B  - 2  \, ,
\end{equation}
whose solution is given by
\begin{equation}
\label{ES5n}
B= 1 + \frac{B_0}{r} + B_1 r^2 \, .
\end{equation}
Here $B_0$ and $B_1$ are constants.
Of course, the solution (\ref{ES5n}) expresses the Schwarzschild-(anti-)de Sitter space-time.

We may also consider the case $F_0=0$.
Then Eq.~(\ref{ES3n})  becomes,
\begin{equation}
\label{ES6n}
0 = - r^2 B'' - r B' + 4 B - 2 \, .
\end{equation}
The solution is given by
\begin{equation}
\label{ES7n}
B= \frac{1}{2} + {\tilde B}_0 r^2 + {\tilde B}_1 r^{-2} \, .
\end{equation}
Here ${\tilde B}_0$ and ${\tilde B}_1$ are constants.
The solution (\ref{ES7n}) corresponds to the solution given before in \cite{Nashed:2019tuk,Elizalde:2020icc}.

In case that either of $F_0$ and $F_1$ does not vanish, when $r$ is small, $F_0$ term in (\ref{ES3n})
dominates and the solution should behave as in (\ref{ES5n}).
On the other hand when $r$ is large, $F_1$ term in (\ref{ES3n}) dominates and the solution should behaves as (\ref{ES7n}).
Then there should exist a solution connect the solution in (\ref{ES5n}) in the small $r$ region and
the solution (\ref{ES7n}) in the large $r$ region.

We may consider a more general case.
By assuming $B\neq 0$, again, Eq.~(\ref{E4n}) can be rewritten as
\begin{equation}
\label{E7n}
N = \exp \left( \int dr \frac{2r F''}{rF' + 2 F} \right) \, .
\end{equation}
We now rewrite Eq.~(\ref{E1n}) as follows,
\begin{align}
\label{E8n}
0=& - 2 r^2 N^2 FB'' + r^2 \left( - 3NN' F - 2N^2F' \right) B' \nn
& + \left[ r^2 \left( F{N'}^2 - 2NN'' F - 3NN'F' + 2 N^2F''\right)
 - 4rN \left( FN' - NF' \right) + 4N^2F \right] B - 4N^2F \, .
\end{align}
By substituting (\ref{E7n}), we obtain,
\begin{align}
\label{E9n}
0=& - 2 r^2 FB'' + r^2 \left( - \frac{6r F F''}{rF' + 2 F} - 2 F' \right) B' \nn
& + \left[ 4F  + 4 r F' - 4 r^2 F'' -2 r^3 F'''
+ \frac{ 6 r^3 F' F'' + 2r^4 F' F'''}{rF' + 2 F}
 - \frac{ 6 r^4 {F'}^2 F''}{\left( rF' + 2 F \right)^2 }
\right] B - 4F \, .
\end{align}
Equation (\ref{E9n}) is the linear inhomogeneous differential equation for $B$ when $F$ is given.
For example, we consider the case $F\propto r^n$ with a constant $n$.
Then Eq.~(\ref{E9n}) reduces
\begin{equation}
\label{E99n}
0=   - 2 r^2 B'' + r \frac{n(-8 n + 6) }{n + 2} B'
+ \frac{8 \left( - n^4 + n^3 + 3 n^2 + 4 n + 2 \right)}{\left( n + 2 \right)^2} B - 4 \, ,
\end{equation}
Let the solution of the following algebraic equation for a constant $\alpha$ be $\alpha_\pm$,
\begin{equation}
\label{E10n}
0 = - 2 \alpha \left( \alpha - 1 \right)  + \frac{n \left( -8 n + 6 \right) }{n + 2} \alpha
+ \frac{8 \left( - n^4 + n^3 + 3 n^2 + 4 n + 2 \right)}{\left( n + 2 \right)^2} \, ,
\end{equation}
that is,
\begin{equation}
\alpha_\pm \equiv \frac{- 2n^2 + 2n + 1 \mp \sqrt{  -4n^3 + 12 n^2 + 20n + 9}}{n+2} \, .
\end{equation}
Then the solution of Eq.~(\ref{E99n}) is given by
\begin{equation}
\label{E11}
B = B_+ r^{\alpha_+} + B_- r^{\alpha_-} + C \, .
\end{equation}
Here $C$ is a constant given by
\begin{equation}
\label{E11}
C = \frac{\left( n + 2 \right)^2}{2 \left( - n^4 + n^3 + 3 n^2 + 4 n + 2 \right)}\, .
\end{equation}
If $C=1$, we have two real solutions $n=0$ and $n=2.39356\cdots$, 	
and two complex solutions $n=-0.696781 \cdots \pm i 0.591668 \cdots$.
The solution $n=0$ gives $\alpha_\pm =-1$, $2$ and it corresponds to
the Schwarzschild-(anti-)de Sitter space-time
in (\ref{ES5n}) but other cases correspond to new kinds of spherically symmetric solutions.

\subsection{New BH}
As we discussed above we have three unknowns in two independent differential equations. Thus to be able to solve these differential equations we assume in this study the unknown function  $F$ to has the form
\begin{eqnarray} \label{ass1n}
F=1+\frac{c_1}{r^2}\,.
\end{eqnarray}
Equation (\ref{ass1n}) shows that when $c_1=0$ we return to the case of GR since in that case $f(R)=\mathrm{cons}.$
\begin{eqnarray} \label{ass1}
&&B(r)=\frac{e^{\frac{3c_1}{2r^2}}}{r}\Bigg\{\mathbb{H}c_2+\mathbb{H}_1r^3c_3+2\mathbb{H}_1r^3\int\frac{e^{\frac{-3c_1}{2r^2}}\mathbb{H}}{r[(2c_1\mathbb{H}_2-
3r^2\mathbb{H})\mathbb{H}_1-2c_1\mathbb{H}\mathbb{H}_3]}dr-2\mathbb{H}\int\frac{e^{\frac{-3c_1}{2r^2}}r^2 \mathbb{H}_1}{(2c_1\mathbb{H}_2-
3r^2\mathbb{H})\mathbb{H}_1-2c_1\mathbb{H}\mathbb{H}_3}dr\Bigg\}\,,\nonumber\\
 &&N(r)=ce^{\frac{-3c_1}{r^2}}\,,\qquad \qquad \qquad \qquad A(r)=N(r)B(r)\,,\qquad \qquad \qquad \qquad F=1+\frac{c_1}{r^2}\,,
\end{eqnarray}
where $\mathbb{H}=\mathrm{HeunC}(\frac{3}{2},\frac{3}{2},0,\frac{3}{8},\frac{9}{8},-\frac{c_1}{r^2})$,  $\mathbb{H}_1=\mathrm{HeunC}(\frac{3}{2},-\frac{3}{2},0,\frac{3}{8},\frac{9}{8},-\frac{c_1}{r^2})$, $\mathbb{H}_2=\mathrm{HeunCPrime}(\frac{3}{2},\frac{3}{2},0,\frac{3}{8},\frac{9}{8},-\frac{c_1}{r^2})$,
$\mathbb{H}_3=\mathrm{HeunCPrime}(\frac{3}{2},-\frac{3}{2},0,\frac{3}{8},\frac{9}{8},-\frac{c_1}{r^2})$\footnote{The $\mathrm{HeunC}$ function is the solution of the Heun Confluent equation which is defined as
\begin{eqnarray} \label{sp1}
X''(r)-\frac{1+\beta-(\alpha-\beta-\gamma-2)r-r^2\alpha}{r(r-1)}X'(r)-\frac{\alpha(1+\beta)-\gamma-2\eta-(1+\gamma)\beta-r(2\delta+[2+\gamma+\beta])}
{2r(r-1)}X(r)=0\,.
\end{eqnarray}
The solution of the above differential equation defined $\mathrm{HeunC}(\alpha,\beta,\gamma,\delta,\eta,r)$ for more details, interested readers can check  \cite{RONVEAUX2003177,MAIER2005171}. The $\mathrm{HeunCPrime}$ is the derivative of the Heun Confluent function.}.
Using Eq. (\ref{ass1}) in the trace equation, i.e., the fourth equation of Eq. (\ref{feq}) we get $f(r)$ in the form
  \begin{eqnarray} \label{ass11}
&& f(r)=-\frac{2e^{^{\frac{3c_1}{2r^2}}}}{r^7}\Bigg\{2r^3\Bigg(3r^2[r^2+c_1]\mathbb{H}_1+2c_1[r^2+3c_1]\mathbb{H}_3\Bigg)\int\frac{e^{^{\frac{-3c_1}{2r^2}}}\mathbb{H}}{r[(2c_1\mathbb{H}_2-
3r^2\mathbb{H})\mathbb{H}_1-2c_1\mathbb{H}\mathbb{H}_3]}dr \nonumber\\
 &&+4c_1\Bigg(3r^2\mathbb{H}-[3c_1+r^2]\mathbb{H}_2\Bigg)\int\frac{e^{^{\frac{-3c_1}{2r^2}}}\mathbb{H}_1r^2}{(2c_1\mathbb{H}_2-
3r^2\mathbb{H})\mathbb{H}_1-2c_1\mathbb{H}\mathbb{H}_3}dr+3c_3r^5[r^2+c_1]\mathbb{H}_1\nonumber\\
 &&+2c_1c_3r^3[r^2+3c_1]\mathbb{H}_3+2c_1c_2[r^2+3c_1]\mathbb{H}_2-6c_1c_2r^2\mathbb{H}-r^3e^{^{\frac{-3c_1}{2r^2}}}[r^2+c_1] \Bigg\},
 \end{eqnarray}
 where $c$, $c_1$, $c_2$ and $c_3$ are constants.
Using Eq. (\ref{ass11}) in  Eq. (\ref{Ricci}) we get
\begin{eqnarray} \label{sol11}
&&R=-\frac{2e^{^{\frac{3c_1}{2r^2}}}}{r^5(r^2+c_1)}\Bigg\{2r^3\Bigg(3r^2[2r^2+c_1]\mathbb{H}_1+2c_1[2r^2+3c_1]\mathbb{H}_3\Bigg)\int\frac{e^{^{\frac{-3c_1}{2r^2}}}\mathbb{H}}{r[(2c_1\mathbb{H}_2-
3r^2\mathbb{H})\mathbb{H}_1-2c_1\mathbb{H}\mathbb{H}_3]}dr \nonumber\\
 &&+4c_1\Bigg(3r^2\mathbb{H}-[3c_1+2r^2]\mathbb{H}_2\Bigg)\int\frac{e^{^{\frac{-3c_1}{2r^2}}}\mathbb{H}_1r^2}{(2c_1\mathbb{H}_2-
3r^2\mathbb{H})\mathbb{H}_1-2c_1\mathbb{H}\mathbb{H}_3}dr+3c_3r^5[2r^2+c_1]\mathbb{H}_1\nonumber\\
 &&+2c_1c_3r^3[2r^2+3c_1]\mathbb{H}_3+2c_1c_2[2r^2+3c_1]\mathbb{H}_2-6c_1c_2r^2\mathbb{H}-2r^3e^{^{\frac{-3c_1}{2r^2}}}[r^2+c_1] \Bigg\}.
\end{eqnarray}
Equations (\ref{ass1}), (\ref{ass11}) and (\ref{sol11}) show that when $c_1=0$ we get
\begin{eqnarray}\label{reda}
 N(r)=c, \qquad \qquad A(r)=B(r) \qquad \qquad \mathrm{ and} \qquad \qquad F(r)=1.\end{eqnarray} Equation (\ref{reda}) shows that when $F(r)=1$ this gives $f(R)=R$ and in that case $A(r)=B(r)=1+\frac{C}{r}$ provided that $c_3=0$. All the above data ensure that when $c_1=0$ we return to the GR BHs\footnote{Note that when $c_1=0$ we get $\mathbb{H}=\mathbb{H}_1=\mathrm{HeunC}(\frac{3}{2},\frac{3}{2},0,\frac{3}{8},\frac{9}{8},0)=\mathrm{HeunC}(\frac{3}{2},
 -\frac{3}{2},0,\frac{3}{8},\frac{9}{8},0)=1$, \\ $\mathbb{H}_2=\mathrm{HeunCPrime}(\frac{3}{2},\frac{3}{2},0,\frac{3}{8},\frac{9}{8},-\frac{c_1}{r^2})=0$ and $\mathbb{H}_3=\mathrm{HeunCPrime}(\frac{3}{2},-\frac{3}{2},0,\frac{3}{8},\frac{9}{8},-\frac{c_1}{r^2})=-\frac{3}{2}$ \cite{RONVEAUX2003177,MAIER2005171}.}.
\subsection{Physical properties of the BH (\ref{ass1})}
We are going, in this section, to understand the physical properties of solution  (\ref{ass1}). For such  aim, we  write the asymptote behaviors of the metric potentials, $A(r)$ and $B(r)$,  given by Eq. (\ref{ass1}) and get
 \begin{eqnarray}\label{mpab}
&& A(r)\approx 1-\frac{2M}{r}-\frac{b^2}{r^2}+\frac{8b^4}{5r^3}+\frac{6b^2M}{r^3}-\frac{52b^4}{21r^4}+\cdots\,,\nonumber\\
 &&
B(r)\approx 1-\frac{2M}{r}+\frac{5b^2}{r^2}+\frac{8b^4}{5r^3}-\frac{6b^2M}{r^3}+\frac{200b^4}{21r^4}+\cdots\,,\end{eqnarray}
where we have assumed $c=1$, $c_1=2b^2$, $c_2=-2M$ and $c_3=0$. Using Eq. (\ref{mpab}) in (\ref{met12}) we get
\begin{eqnarray} \label{metaf}
& &  ds^2=-\Bigg[1-\frac{2M}{r}-\frac{b^2}{r^2}+\frac{8b^4}{5r^3}+\frac{6b^2M}{r^3}-\displaystyle\frac{52b^4}{21r^4}\Bigg]dt^2+\frac{dr^2}{ 1-\frac{2M}{r}+\frac{5b^2}{r^2}+\frac{8b^4}{5r^3}-\frac{6b^2M}{r^3}+\frac{200b^4}{21r^4}}+r^2(d\theta^2+\sin^2d\phi^2)\,. \nonumber\\
 && \end{eqnarray}
The line element (\ref{metaf}) is asymptotically approaching a flat spacetime and does not coincide with the Schwarzschild spacetime due to the contribution of the extra terms that come mainly from the constant parameter $b$ whose source is the effect of higher-order curvature terms of  $\mathit{f(R)}$. As one can check easily that when these extra terms equal zero one can smoothly return to the Schwarzschild spacetime \cite{Misner:1974qy}. We assume the constant $c_3=0$ in the asymptote of the metric potential (\ref{mpab}). This assumption makes the metric potentials asymptote to flat spacetime. When the constant $c_3\neq 0$ the metric potentials will not asymptote to flat spacetime but to AdS/dS spacetime as follows:
\begin{eqnarray}\label{mpab1}
&& A(r)\approx \pm\Lambda_\mathrm{eff} r^2+1-\frac{2M}{r}-\frac{b^2}{r^2}+\frac{8b^4}{5r^3}+\frac{6b^2M}{r^3}-\frac{52b^4}{21r^4}+\cdots\,,\nonumber\\
 &&
B(r)\approx  \pm\Lambda_\mathrm{eff}r^2+1\pm2-\frac{2M}{r}+\frac{(5\pm6)b^2}{r^2}+\frac{8b^4}{5r^3}-\frac{6b^2M}{r^3}+\frac{(200\pm252)b^4}{21r^4}+\cdots\,,\end{eqnarray}
where $\Lambda_\mathrm{eff}=c_3=\pm\frac{1}{3b^2}$.
Using Eq. (\ref{mpab1}) in (\ref{met12}) we get
\begin{eqnarray} \label{metaf1}
& &  ds^2=-\Bigg[\pm\Lambda_\mathrm{eff} r^2+1-\frac{2M}{r}-\frac{b^2}{r^2}+\frac{8b^4}{5r^3}+\frac{6b^2M}{r^3}-\frac{52b^4}{21r^4}\Bigg]dt^2\nonumber\\
 &&+\displaystyle\frac{dr^2}{  \pm\Lambda_\mathrm{eff}+1\pm2-\frac{2M}{r}+\frac{(5+\pm6)b^2}{r^2}+\frac{8b^4}{5r^3}-\frac{6b^2M}{r^3}+\frac{(200\pm252)b^4}{21r^4}}+r^2(d\theta^2+\sin^2d\phi^2)\,. \nonumber\\
 && \end{eqnarray}
The line element (\ref{metaf}) is asymptotically approaches AdS/dS spacetime  according to the sign of $\Lambda_\mathrm{eff}$.

Now we are going to use Eq. (\ref{metaf1})  in Eq. (\ref{Ricci})  and get
\begin{eqnarray} \label{R1}
&&R(r)\approx -12c_3-\frac{18c_1c_3}{r^2}+\frac{6c_1}{r^4}+\cdots\equiv-12\Lambda_\mathrm{eff}\mp\frac{12}{r^2}+\frac{12b^2}{r^4}+\cdots\,,\nonumber\\
 &&r(R)=\pm\frac{\sqrt{2(\mp3\pm\sqrt{9\pm12+3R})}}{\sqrt{12\Lambda_\mathrm{eff}+R}}\approx \pm0.88954361b\mp0.019200037b^3R\pm0.0013073405b^5R^2\mp0.0001157435b^7 R^3+\cdots \,,\nonumber\\
 &&
\end{eqnarray}
where we have put $c_3=\frac{1}{3b^2}$ which correspond to AdS spacetime and the other two roots are neglected because they give imaginary quantities.
Equation (\ref{R1}) shows that when the constant $c_3=0$ we have a non-trivial value of the Ricci scalar which contributes to higher-order curvature and when $c_1=0$ we get a trivial value of the Ricci scalar which corresponds to GR BH. The asymptote form of  $f(r)$,  given by Eq. (\ref{ass11}), has the form
\begin{eqnarray} \label{fR1}
&&f(r)\approx -6\Lambda_\mathrm{eff}-\frac{16}{r^2}-\frac{16b^2}{r^3}-\frac{50b^2}{r^4}-\frac{201.6b^4}{r^5}-\frac{48b^2M}{r^5}-\frac{60b^4}{r^6}+\cdots\,, \quad {\textrm for\, \, c_3=\frac{1}{3b^2}}\,,\nonumber\\
 &&\approx 6\Lambda_\mathrm{eff}+\frac{16}{r^2}+\frac{16b^2}{r^3}+\frac{70b^2}{r^4}+\frac{182b^4}{r^5}-\frac{48b^2M}{r^5}+\frac{228b^4}{r^6}+\cdots\,, \quad {\textrm for\, \, c_3=\frac{-1}{3b^2}}\,.\nonumber\\
\end{eqnarray}
Using second equation of (\ref{R1}) in (\ref{fR1}) we get
\begin{eqnarray} \label{fR2}
&&f(R)\approx C_1+C_2R+C_3R^2+C_4R^3\,,\nonumber\\
 \end{eqnarray}
 with  $C_i, i=1\cdots 4$ are constants that their values depend on the sign of $c_3\pm\frac{1}{3b^2}$\footnote{The constants $C_i, i=1\cdots 4$ have different values depending on the sign of $c_3=\pm\frac{1}{3b^2}$.  For example when  $c_3=\frac{1}{3b^2}$,  the constants $C_i$ take the values $C=-(669.5280519\Lambda_\mathrm{eff}\pm\frac{384.6857554}{b}\pm\frac{86.17970730M}{b^3})$, $C_1=-(23.45043154\pm40.53430523b\pm\frac{9.300575885M}{b})$, $C_2=(\pm0.1670765328b^3\pm0.03104615112qM+0.01169075762b^2)$ and  $C_3=-(\pm0.02091872050b^5\pm0.004384004185b^3M+0.01057683849b^4)$.}.

 Now use Eq.  (\ref{ass1}) in order to calculate the invariants  to obtain
 \begin{eqnarray} \label{inv}
&& R_{\mu \nu \rho \sigma} R^{\mu \nu \rho \sigma}= 24\Lambda_\mathrm{eff}{}^2+\frac{16}{b^2r^2}+\frac{32}{r^4}+\frac{64M}{r^5}+\frac{64b^2}{15r^5}\cdots\,, \nonumber\\
 &&
  R_{\mu \nu } R^{\mu \nu }=36\Lambda_\mathrm{eff}{}^2+\frac{24}{b^2r^2}+\frac{24}{r^4}+\frac{96M}{r^5}+\frac{32b^2}{5r^5}\cdots\,, \nonumber\\
 &&R=-12\Lambda_\mathrm{eff}-\frac{12}{r^2}+\frac{12b^2}{r^4}-\frac{48b^2M}{r^5}-\frac{16b^4}{5r^5}\cdots,
 \end{eqnarray}
  with $\Big(R_{\mu \nu \rho \sigma} R^{\mu \nu \rho \sigma}, R_{\mu \nu} R^{\mu \nu}, R\Big)$ are the Kretschmann
scalar, the Ricci tensor square, the Ricci scalar, respectively and all of them have a true singularity at $r=0$. Moreover, the above equations show that $b^2=c_1/2$ must not equal zero. It is important to stress on the fact that  the constant $c_1$ is the main source for the deviation of the above results from GR that has the following values $\Big(R_{\mu \nu \rho \sigma} R^{\mu \nu \rho \sigma}, R_{\mu \nu} R^{\mu \nu}, R\Big)=(16[2\Lambda^2r^6+9M]/[3r^6],16\Lambda^2,\mp8\Lambda)$.
 Equation (\ref{inv}) indicates that the leading term of the invariants  $(R_{\mu \nu \rho \sigma} R^{\mu \nu \rho \sigma},R_{\mu \nu} R^{\mu \nu},R)$ is $(\frac{1}{r^2},\frac{1}{r^2},\frac{1}{r^2})$ which is different from  the Schwarzschild  BH which gives the leading term of the Kretschmann
scalar as  $\frac{1}{r^\mathrm{6}}$ and the other invariants  $R_{\mu \nu} R^{\mu \nu}=R=\mathrm{const.}$ Therefore, Eq. (\ref{inv}) indicates that  Kretschmann   singularity  is milder  than the Schwarzschild BH of GR.
\section{Thermodynamics of the BH}\label{S4}
We are going in this section to study the properties  the BHs (\ref{mpab}) and (\ref{mpab1}) from the viewpoint of thermodynamics. For this aim, we  will write the basic definitions of the  quantities of thermodynamics that we will use. The surface gravity of a spacetime having two horizons is defined as:
\begin{equation}
\chi_{\pm}=\frac{r_\pm-r_\mp}{2r_\pm{}^2}\,,
\end{equation}
where $r_\pm$ are the inner and outer horizons of the spacetime.
The  temperature of Hawking is given by \cite{PhysRevD.86.024013,Sheykhi:2010zz,Hendi:2010gq,PhysRevD.81.084040,Wang:2018xhw,Zakria:2018gsf}
\begin{equation}\label{temp}
T_\pm = \frac{\chi_{\pm}}{2\pi}=\frac{r_\pm-r_\mp}{4\pi r_\pm{}^2}\,.
\end{equation}
The semi classical Bekenstein-Hawking entropy of the horizons is defined as
\begin{equation}\label{ent}
\delta_\pm =\frac{1}{4}\Big(A_\pm\Big) f_R\,,
\end{equation}
with $A_\pm$  being the area of the horizons.  The quasi-local  energy is figured out as \cite{PhysRevD.84.023515,PhysRevD.86.024013,Sheykhi:2010zz,Hendi:2010gq,PhysRevD.81.084040,Zheng:2018fyn}
\begin{equation}\label{en}
E(r_\pm)=\frac{1}{4}\displaystyle{\int }\Bigg[2f_{R}(r_\pm)+r_\pm{}^2\Big\{f(R(r_\pm))-R(r_\pm)f_{R}(r_\pm)\Big\}\Bigg]dr_\pm.
\end{equation}
Finally,   the Gibbs free energy  is figured out as \cite{Zheng:2018fyn,Kim:2012cma}
\begin{equation} \label{enr}
G(r_\pm)=E(r_\pm)-T(r_\pm)S(r_\pm).
\end{equation}
\subsection{Thermodynamics of solution  (\ref{mpab}) that has a flat spacetime}
\begin{figure}
\centering
\subfigure[~The metric potential of BH (\ref{mpab})]{\label{fig:met}\includegraphics[scale=0.3]{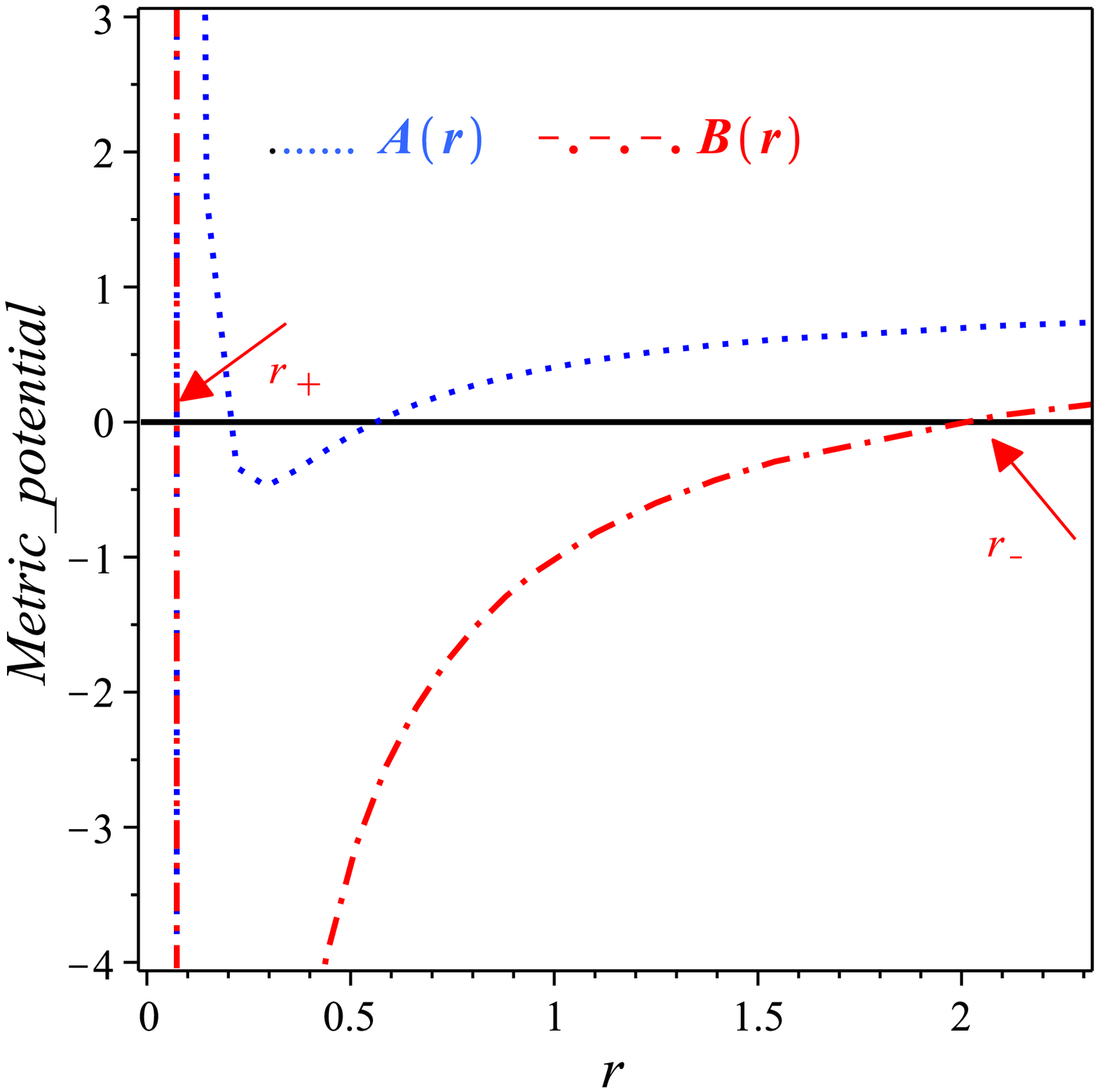}}
\subfigure[~The metric potential of BH (\ref{mpab})]{\label{fig:metrd}\includegraphics[scale=0.3]{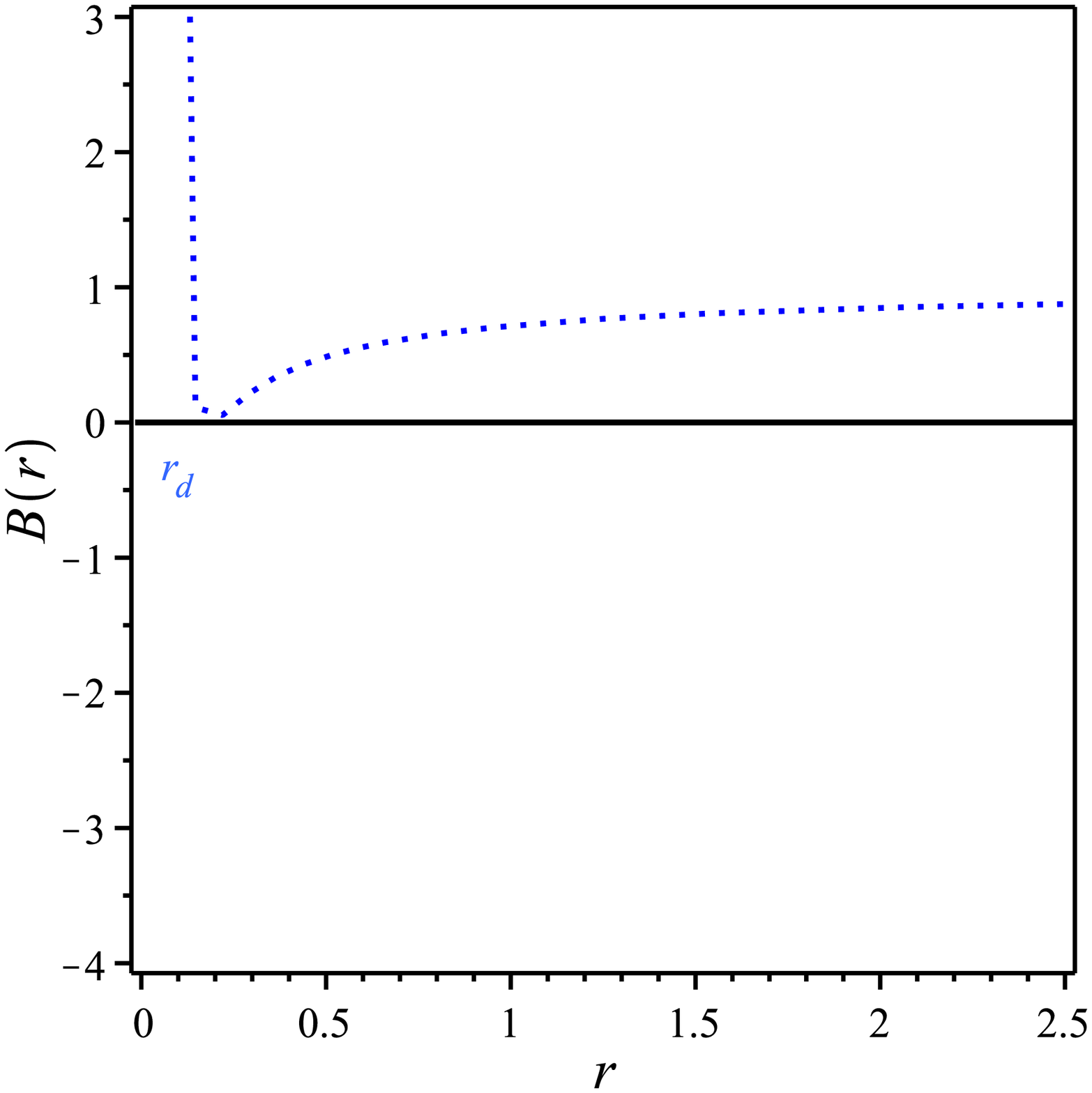}}
\subfigure[~Hawking temperature of BH (\ref{mpab})]{\label{fig:Hor}\includegraphics[scale=0.3]{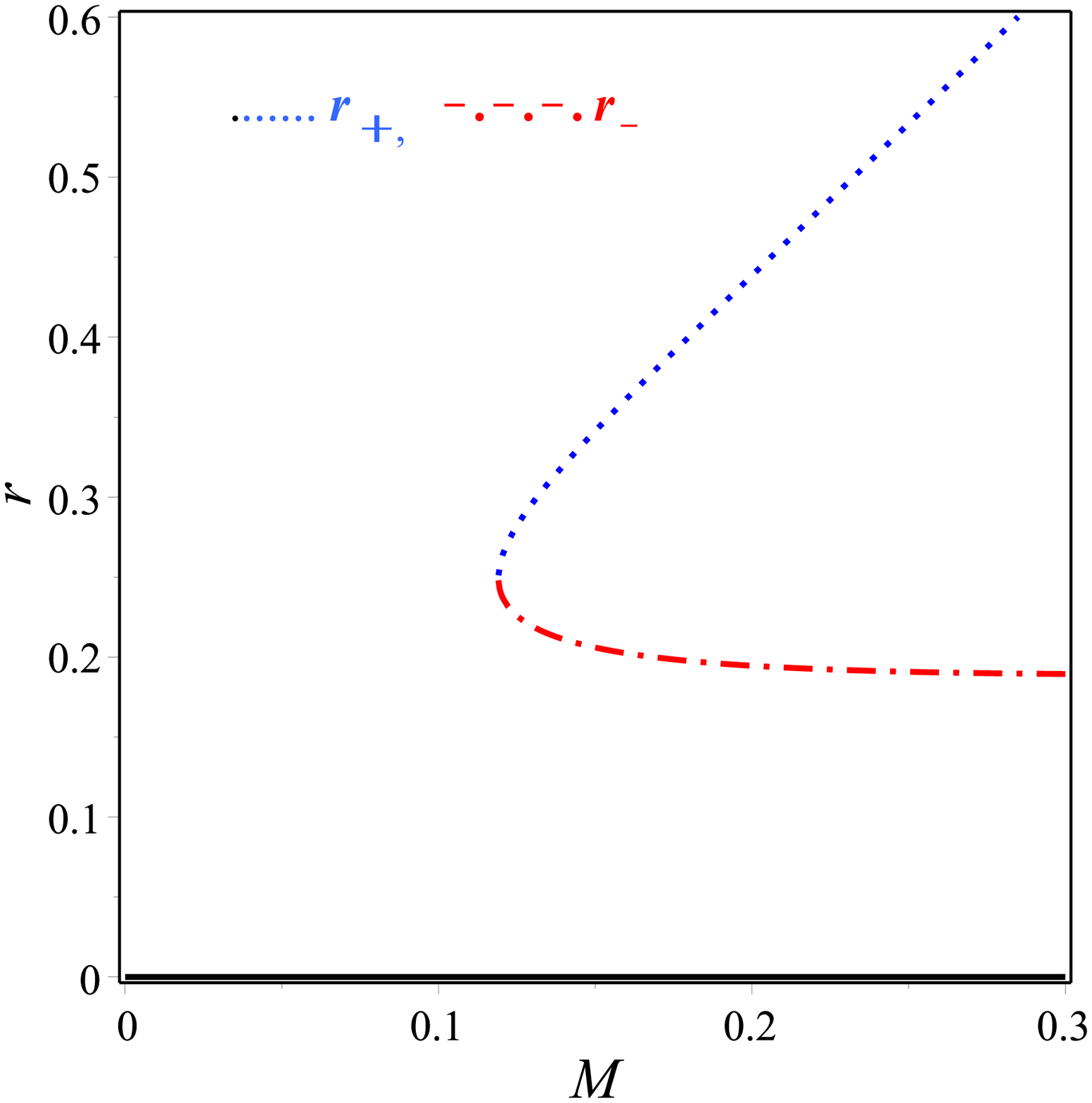}}
\caption[figtopcap]{\small{{Plot of the horizons given by Eq. (\ref{r1}) using $b=0.1$ which is consistent with the constrains (\ref{cons}).}}}
\label{Fig:1}
\end{figure}

\begin{figure}
\centering
\subfigure[~Hawking temperature of BH  (\ref{mpab})]{\label{fig:Temp}\includegraphics[scale=0.4]{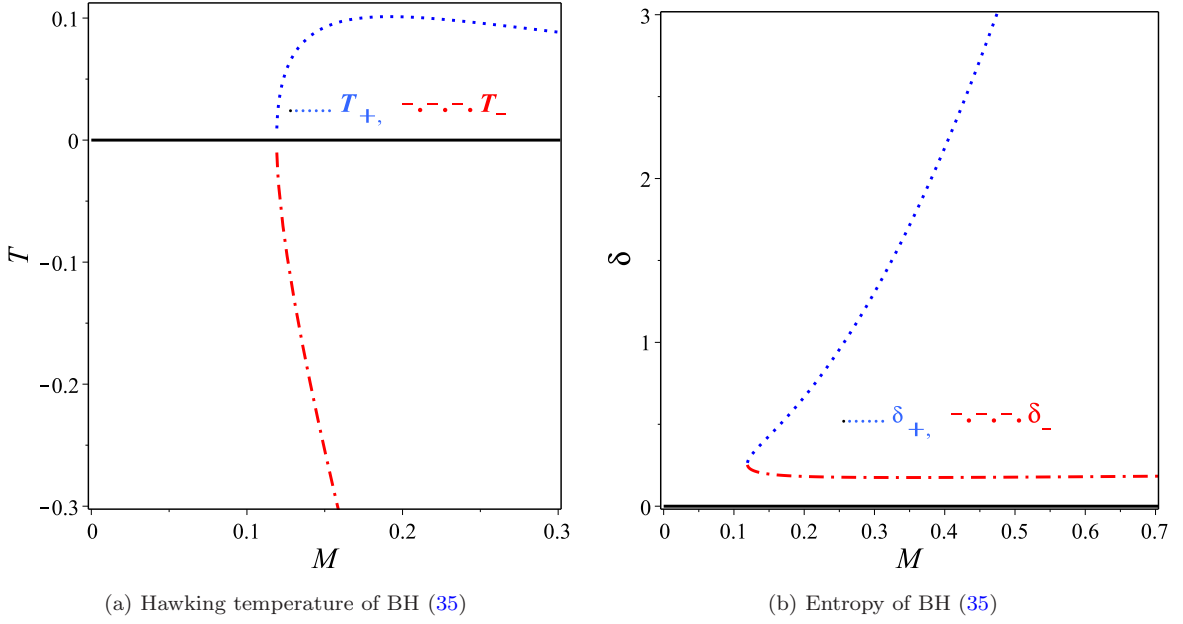}}
\subfigure[~Entropy of BH (\ref{mpab})]{\label{fig:ent}\includegraphics[scale=0.4]{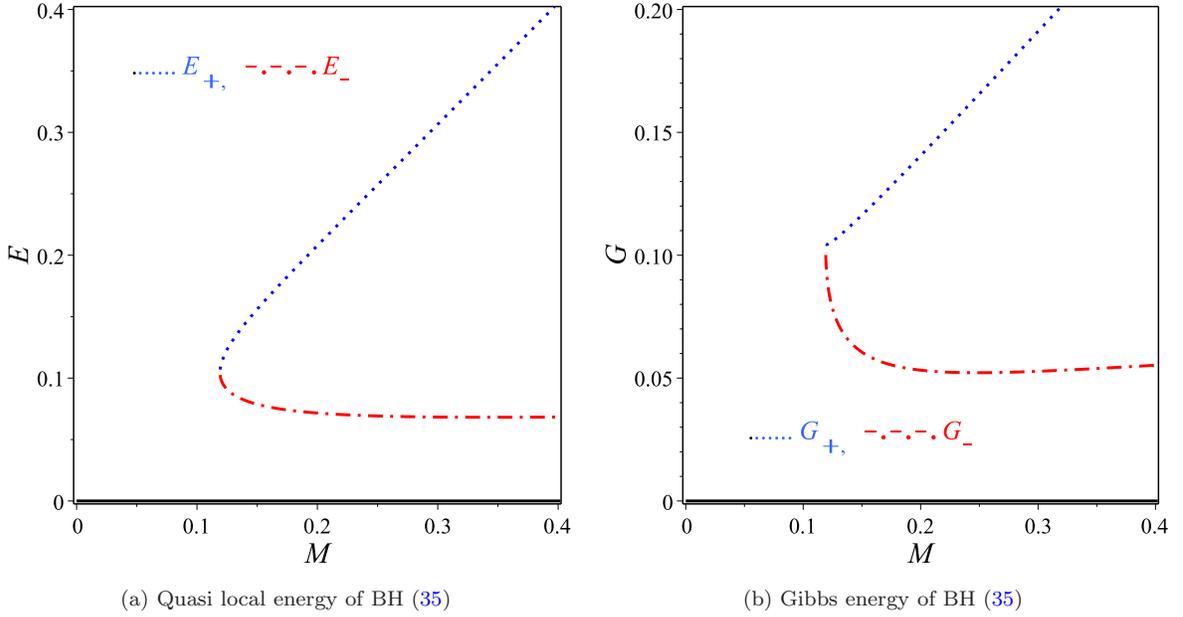}}
\caption[figtopcap]{\small{{Plot of the horizons given by Eq. (\ref{r1}) using $b=0.1$ which is consistent with the constrains (\ref{cons}).}}}
\label{Fig:2}
\end{figure}

The BH (\ref{mpab})   derived in the previous section is portrayed by the mass of the BH $M$ and the parameter $b$ and when the parameter $b$ is vanishing we get the Schwarzschild spacetime which corresponds to GR. To find the horizons of this BH, (\ref{mpab}), we put $A(r)=0$. This gives four roots two of them are real and the others are imaginary. The real roots have the form
\begin{eqnarray}\label{r1}
&&r_\pm=\frac{[14910M^{4/3}+\sqrt{710}X_1{}^2]X_1\pm\sqrt{710}\sqrt{Y_1X_1{}^2+7.008182433\times10^7M(5M^3-5Mb^2-8b^4)}}{29820M^{1/3}X_1}\,,\nonumber\\
 %
\end{eqnarray}
where $X_1=\Bigg[993.4163240(112M+1625)b^{8/3}+313110M^{8/3}+1043700b^2M^{2/3}+9.398818978\times10^5b^{4/3}M^{4/3}+4.172348561\times10^5b^{2/3}M^2\Bigg]^{1/4}$
 and $Y_1=\Bigg[-993.4163240(112M+1625)b^{8/3}+626220M^{8/3}+2087400b^2M^{2/3}-9.398818978\times10^5b^{4/3}M^{4/3}-4.172348561\times10^5b^{2/3}M^2\Bigg]$. Equation (\ref{r1}), $r_\pm$,  put the following constraint to have a real value \begin{eqnarray}\label{cons}
 Y_1X_1{}^2+7.008182433\times10^7(5M^3-5Mq^2-8b^4)>0.\end{eqnarray}
The metric potentials of the BH (\ref{mpab}) are drawn  in Fig. \ref{Fig:1} \subref{fig:met}. From Fig. \ref{Fig:1} \subref{fig:met} we can easy see the  two horizons of the metric potentials $A(r)$ and  $B(r)$.  Also the behavior of the horizons given by Eq. (\ref{r1}) are drawn in Fig. \ref{Fig:1} \subref{fig:Hor}.  It is easy to check that  the degenerate horizon for the metric potential $B(r)$  is happened for a specific value of $(b,M,r)\equiv(0.1,0.163,0.1647332393)$, respectively which correspond to the Nariai BH. The degenerate behavior is shown is Fig. \ref{Fig:1} \subref{fig:metrd}. The Fig. \ref{Fig:1} \subref{fig:Hor} shows that  the horizon $r_+$ increasing with $M$ while $r_-$ decreasing.
Using Eq. (\ref{temp}), the Hawking temperature  can be calculated as:
\begin{eqnarray}\label{T1}
&&T_\pm=\nonumber\\
&& \pm\frac{14058.36M^{1/3}\sqrt{X_1{}^2Y_1+[350409121.7M^4-350409121.7M^2b^2-560654594.7b^4M]}X_1{}^{1/4}}
{\Bigg((4970M^{4/3}+8.88194173X_2{}^{2})X_1\pm8.88\sqrt{Y_1X_1{}^2+350409121.7M^4-350409121.7M^2 b^2-560654594.7b^4M}\Bigg)^2}\,.\nonumber\\
&&
\end{eqnarray}
The behavior of the Hawking temperature   given by Eq. (\ref{T1}) is drawn in Fig. \ref{Fig:2} \subref{fig:Temp} which shows that $T_{+}>T_{-}$. As Fig. \ref{Fig:2} \subref{fig:Temp} shows that the $T_+$ has an increasing positive temperature while $T_-$ has decreasing negative temperature.  Figure \ref{Fig:2} \subref{fig:Temp} indicates that   $T_+$ has a vanishing  value at  $r_+ = r_d$. Moreover, when  $r_+ < r_d$, $T_-$  becomes  negative and  an ultracold BH is formed. Also,  Davies \cite{Davies:1978mf} clarified that there is no clear reason from thermodynamical effects to prevent BH temperature to be below the absolute zero and in that case  a naked singularity is formed.  Figure  \ref{Fig:2} \subref{fig:Temp}  shows  Davies  argument at $r_+ < r_\mathrm{min}$ region.

\begin{figure}
\centering
\subfigure[~Quasi local energy of BH (\ref{mpab})]{\label{fig:Enr}\includegraphics[scale=0.4]{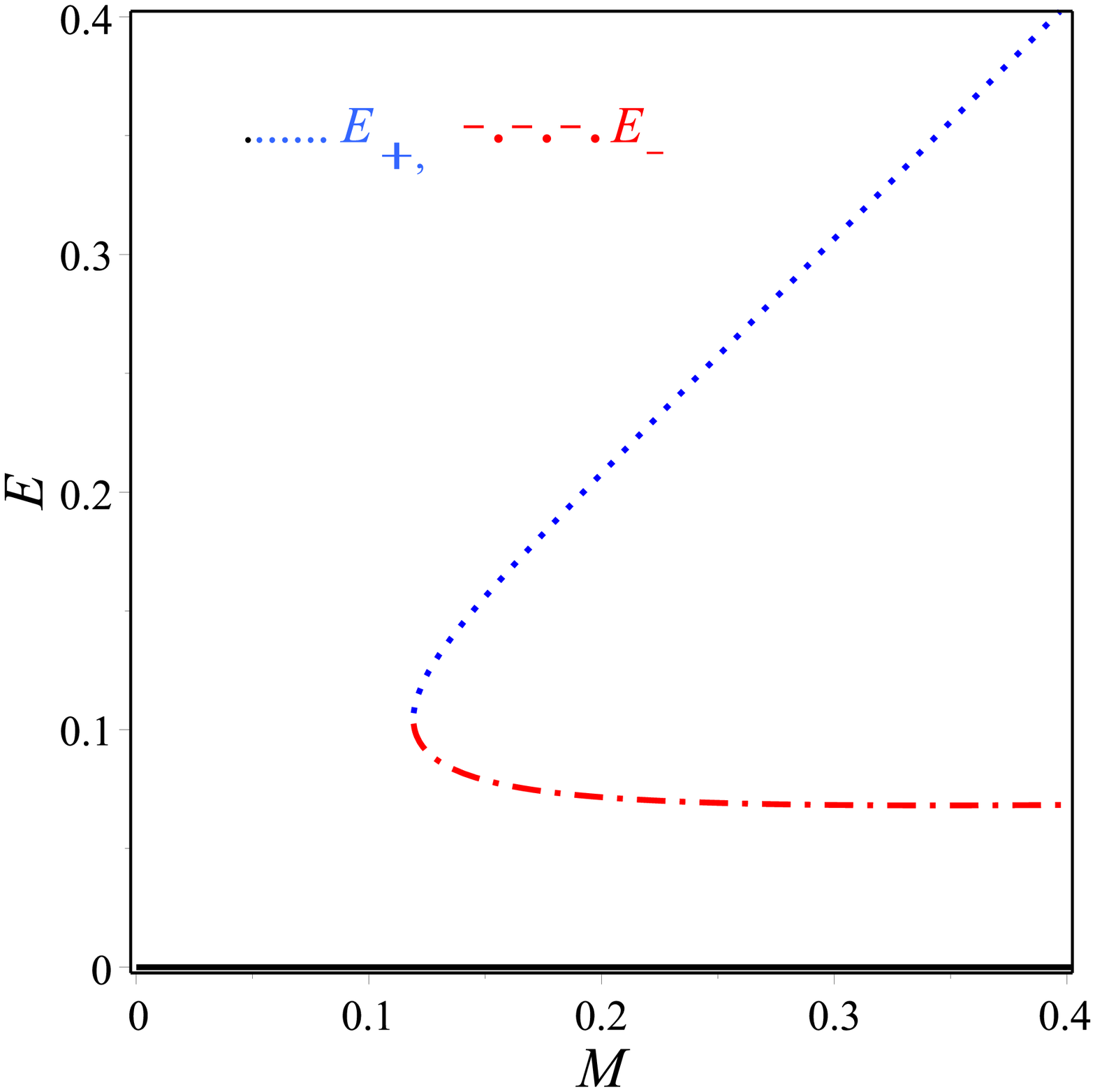}}
\subfigure[~Gibbs energy of BH (\ref{mpab})]{\label{fig:gib}\includegraphics[scale=0.4]{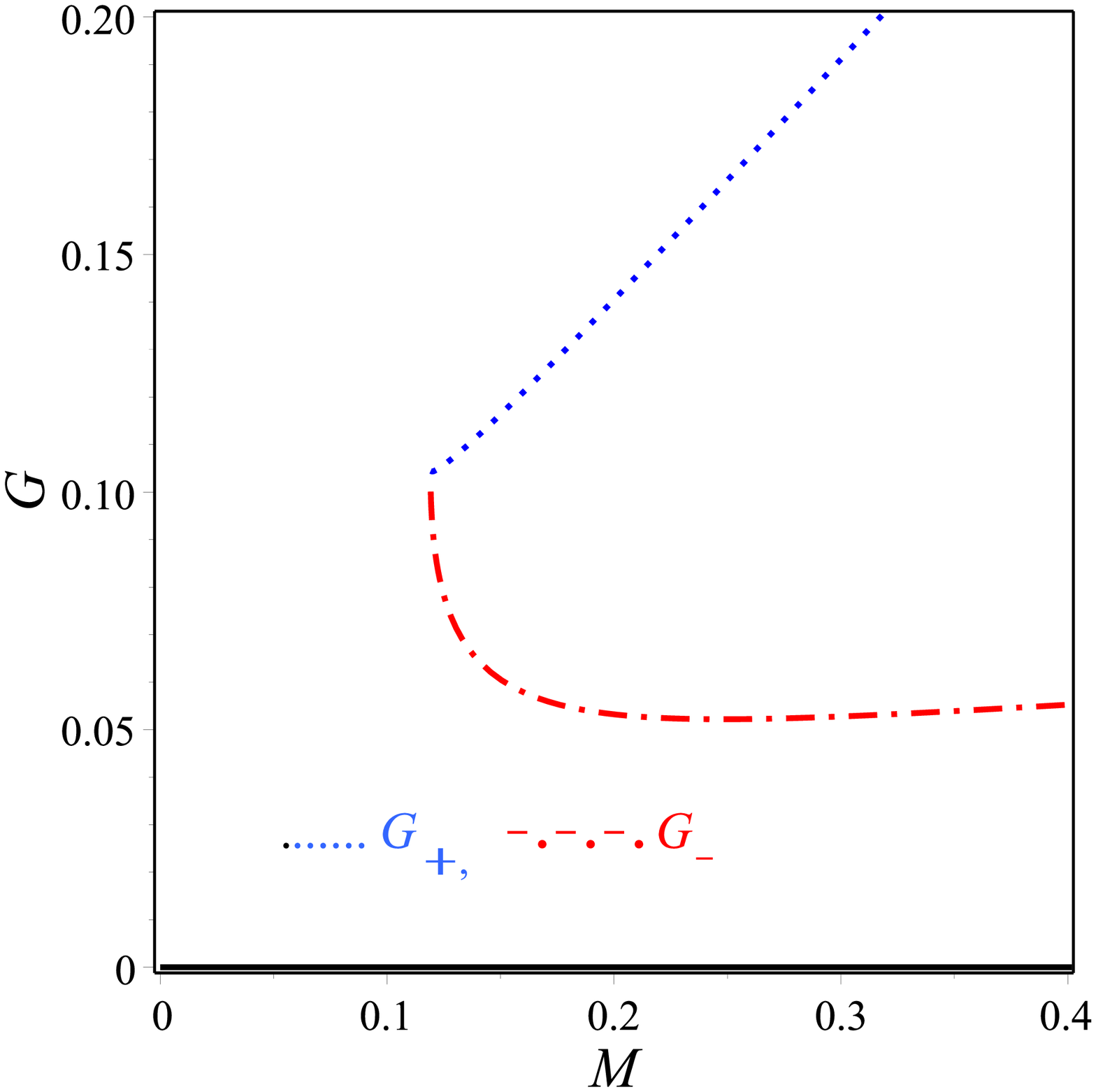}}
\caption[figtopcap]{\small{{Plot of the horizons given by Eq. (\ref{r1}) using $b=0.1$ which is consistent with the constrains (\ref{cons}).}}}
\label{Fig:3}
\end{figure}

Using Eq. (\ref{ent}) we get the entropy of BH  (\ref{mpab}) in the form
\begin{eqnarray}\label{S1}
&&\delta_\pm= \frac{3.531135393\Bigg([14910M^{4/3}+\sqrt{710}X_1{}^2]X_1\pm\sqrt{710}\sqrt{Y_1X_1{}^2+7.008182433\times10^7M(5M^3-5Mq^2-8b^4)}\Bigg)^2}{10^{9}M^{2/3}X_1{}^2}+\pi b^2\,.\nonumber\\
&&
\end{eqnarray}
The   entropy behavior is given in Fig. \ref{Fig:2} \subref{fig:ent} that indicates an increasing value for $\delta_+$ and decreasing value for $\delta_-$. From Eq. (\ref{en}),  the quasi-local energy takes the form
\begin{eqnarray}\label{E1}
&&E_\pm=\frac{[14910M^{4/3}+\sqrt{710}X_1{}^2]X_1\pm\sqrt{710}\sqrt{Y_1X_1{}^2+7.008182433\times10^7M(5M^3-5Mb^2-8b^4)}}{59640M^{1/3}X_1}\nonumber\\
&&-\frac{29820M^{1/3}X_1b^2}{2\{[14910M^{4/3}+\sqrt{710}X_1{}^2]X_1\pm\sqrt{710}\sqrt{Y_1X_1{}^2+7.008182433\times10^7M(5M^3-5Mb^2-8b^4)}\}}\,.
\end{eqnarray}
The quasi-local energies behavior are shown in Fig. \ref{Fig:3} \subref{fig:Enr} which also shows positive increasing value for $E_+$ and positive decreasing value for $E_-$. Finally, we use Eqs. (\ref{T1}),  (\ref{S1}) and  (\ref{E1}) in Eq.  (\ref{enr}) to calculate the Gibbs free energies. The behavior of these free energies are  in Fig. \ref{Fig:3} \subref{fig:gib} which shows positive increasing for $G_+$ and positive decreasing for $G_-$.

\subsection{Thermodynamics of the BH (\ref{mpab1}) that has  asymptote flat AdS}
\begin{figure}
\centering
\subfigure[~The metric potential of BH (\ref{mpabds1})]{\label{fig:met}\includegraphics[scale=0.3]{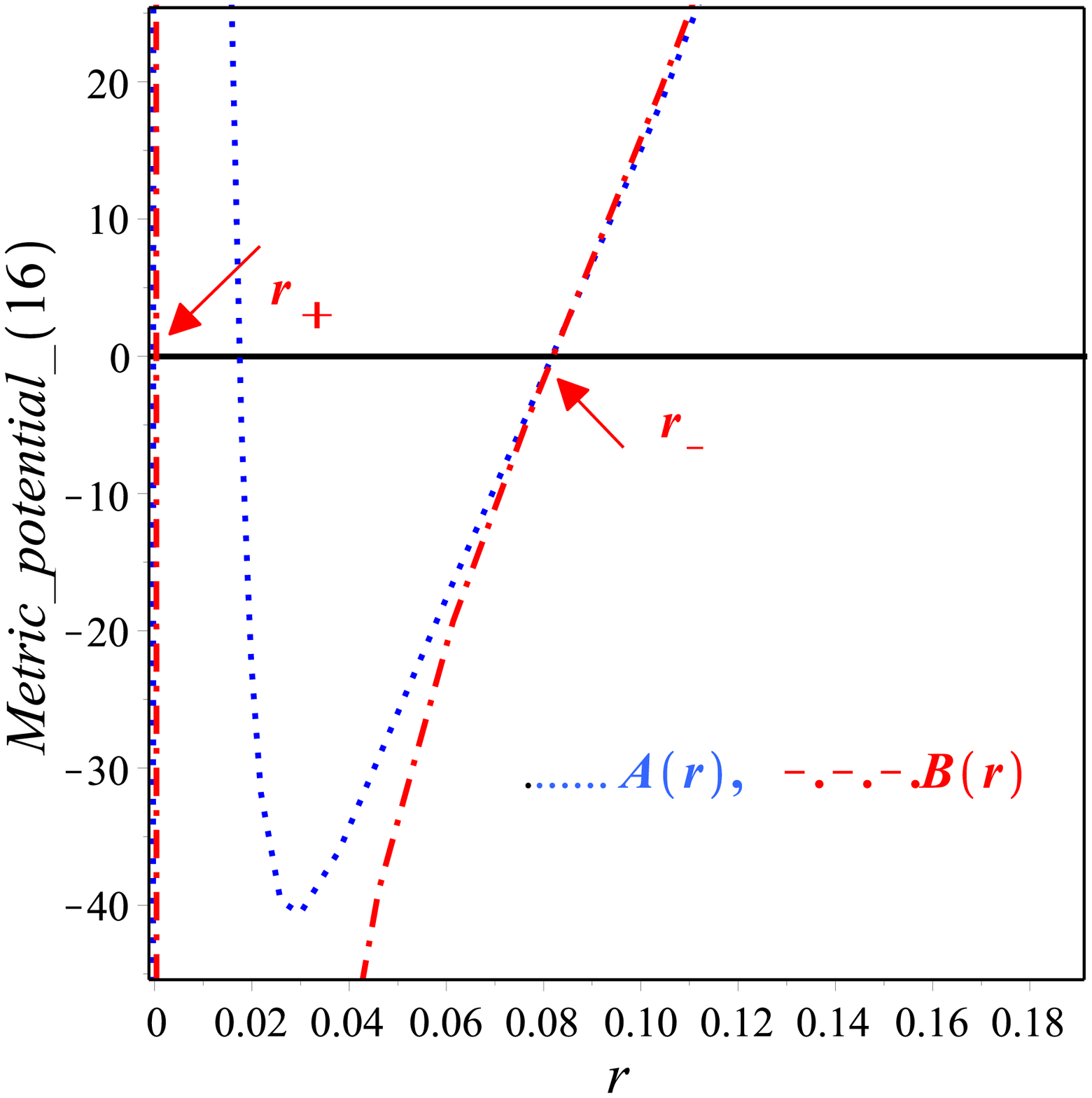}}
\subfigure[~The metric potential of BH (\ref{mpabds1})]{\label{fig:metrd}\includegraphics[scale=0.3]{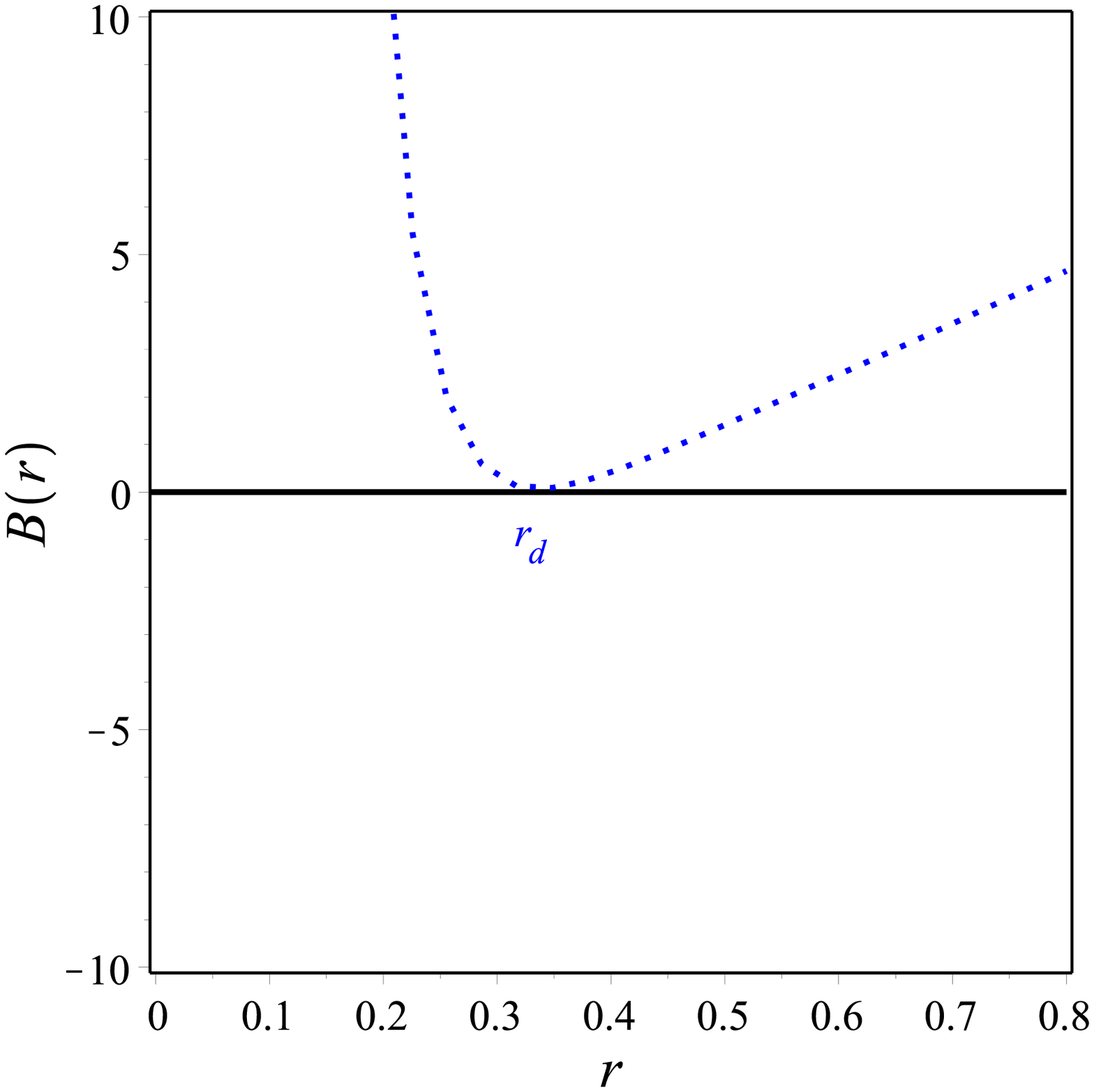}}
\subfigure[~Hawking temperature of BH (\ref{mpabds1})]{\label{fig:Hor}\includegraphics[scale=0.3]{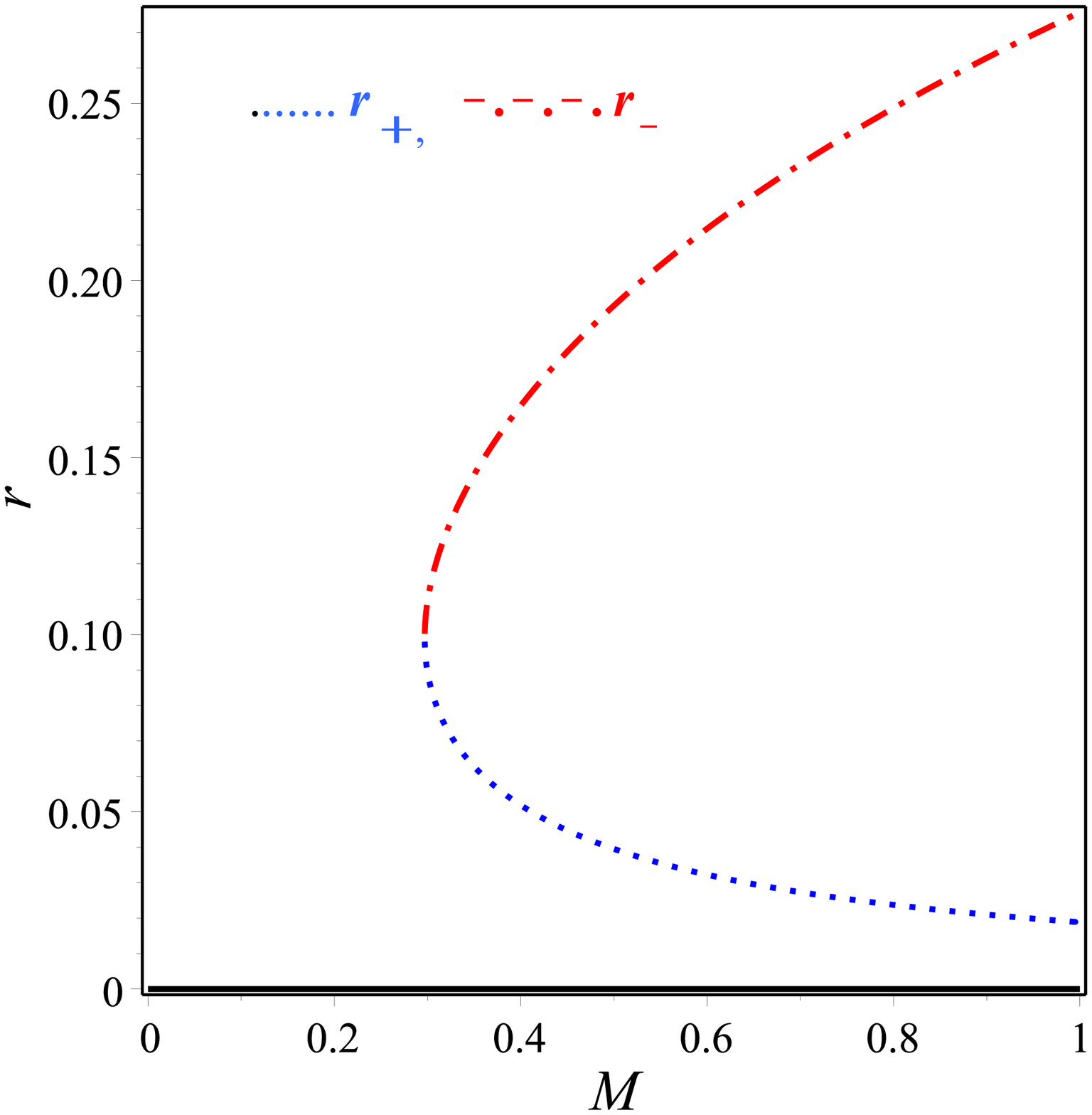}}
\caption[figtopcap]{\small{{Plot of the horizons given by Eq. (\ref{r1}) using $b=0.1$}}}
\label{Fig:4}
\end{figure}

\begin{figure}
\centering
\subfigure[~Hawking temperature of BH  (\ref{mpabds1})]{\label{fig:Temp}\includegraphics[scale=0.4]{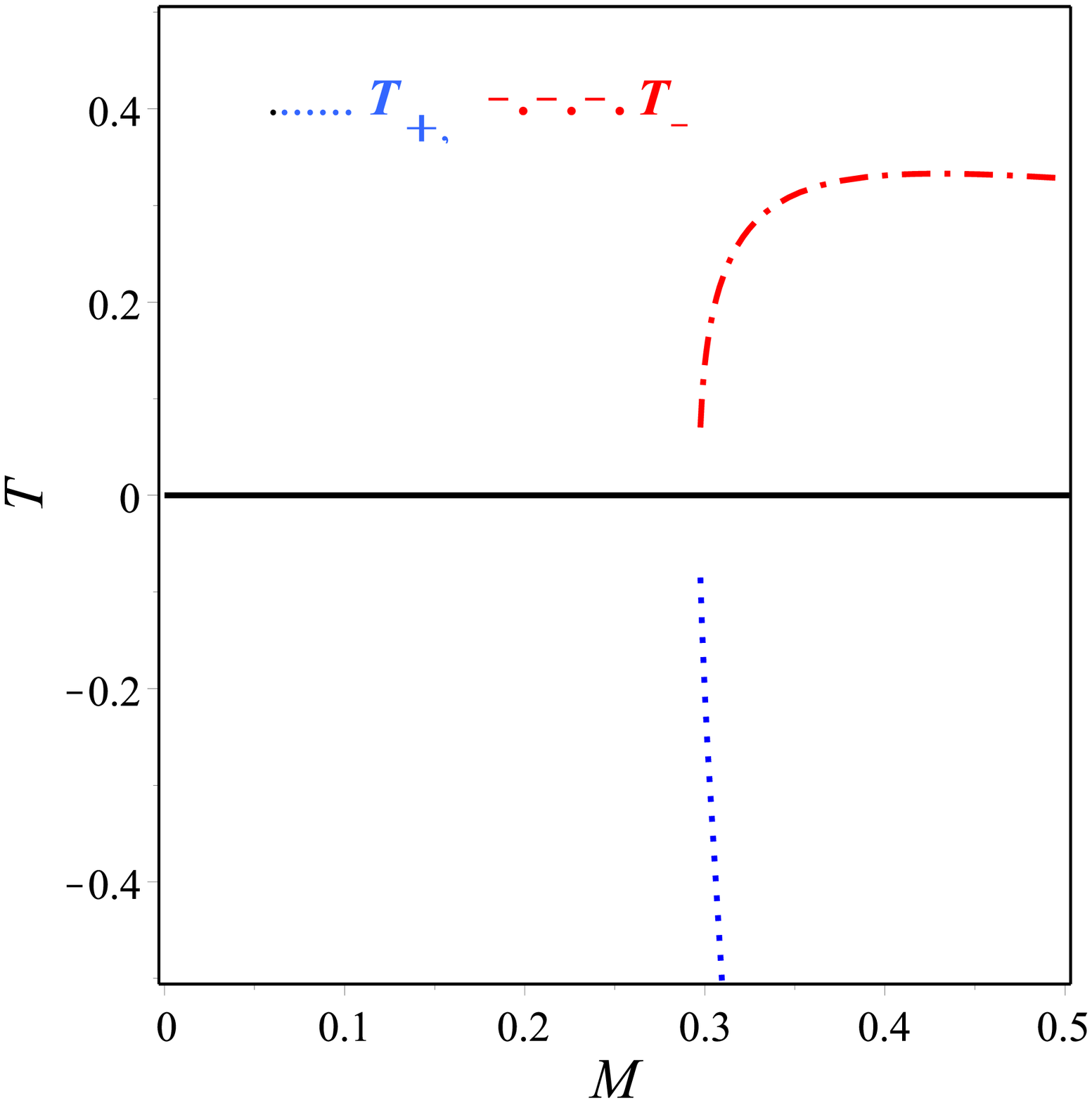}}
\subfigure[~Entropy of BH (\ref{mpabds1})]{\label{fig:ent}\includegraphics[scale=0.4]{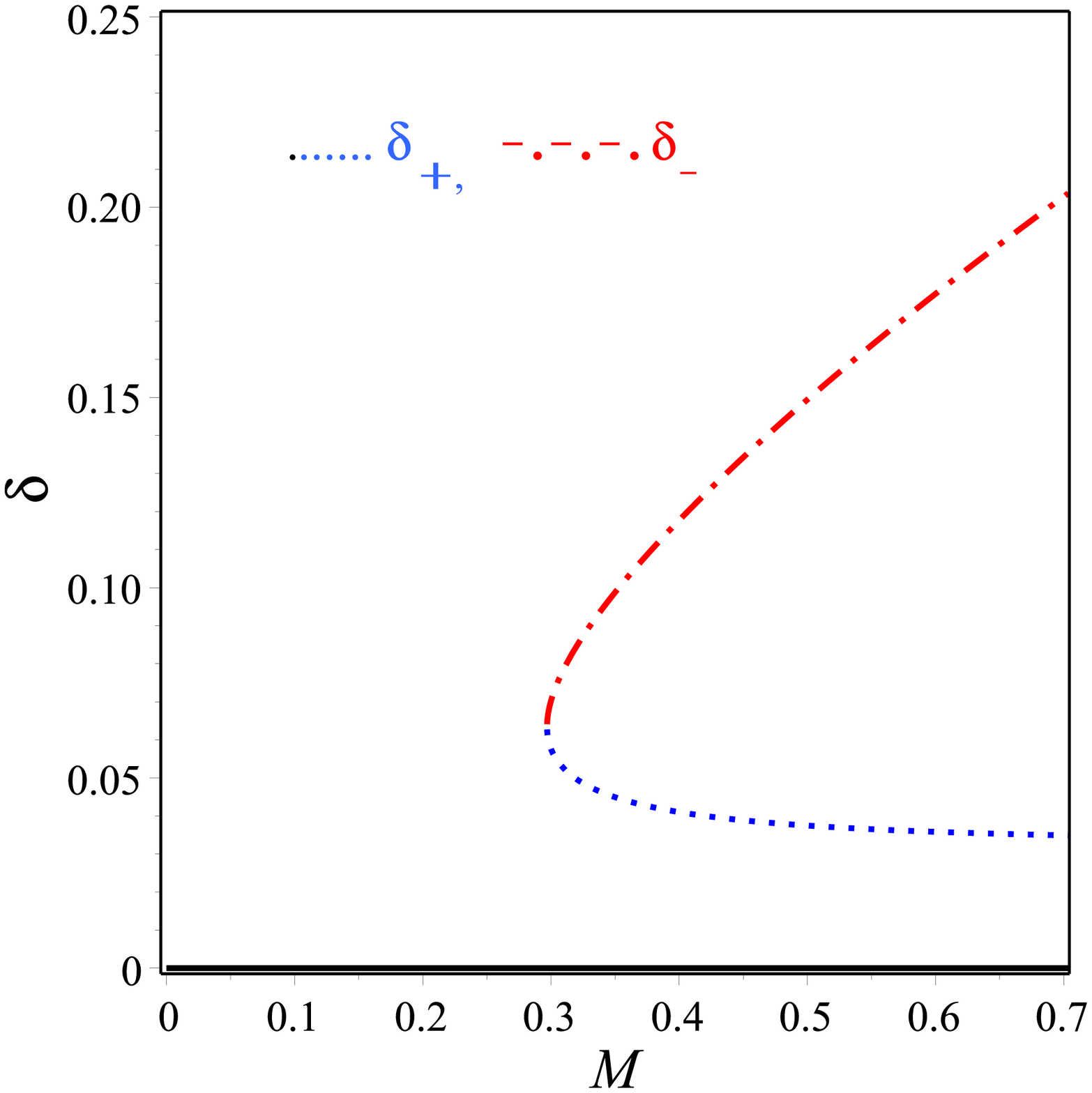}}
\caption[figtopcap]{\small{{Plot of the horizons given by Eq. (\ref{r1}) using $b=0.1$ which is consistent with the constrains (\ref{cons}).}}}
\label{Fig:5}
\end{figure}

In this subsection  we are going to study the BH (\ref{mpab1}) which is characterized by the mass of the BH $M$ and the parameter $b$ and a positive cosmological effective constant. The metric potential when $\Lambda_\mathrm{eff.}=\frac{1}{3b^2}$ takes the form
\begin{eqnarray}\label{mpabds1}
&& A(r)\approx \Lambda_\mathrm{eff} r^2+1-\frac{2M}{r}-\frac{b^2}{r^2}+\frac{8b^4}{5r^3}+\frac{6b^2M}{r^3}-\frac{52b^4}{21r^4}+\cdots\,,\nonumber\\
 &&
B(r)\approx  \Lambda_\mathrm{eff}r^2+3-\frac{2M}{r}+\frac{11b^2}{r^2}+\frac{8b^4}{5r^3}-\frac{6b^2M}{r^3}+\frac{452b^4}{21r^4}+\cdots\,.\end{eqnarray}
When the parameter $b$ is vanishing we get the Schwarzschild AdS spacetime which corresponds to the Einstein GR. The metric potentials of the BH (\ref{mpabds1}) are drawn  in Fig. \ref{Fig:4} \subref{fig:met}. From Fig. \ref{Fig:4} \subref{fig:met} we can easy see the  two horizons of the metric potentials $A(r)$ and  $B(r)$. To find the horizons of this BH, (\ref{mpabds1}), we put $A(r)=0$ in Eq. (\ref{mpabds1}) \cite{Wang:2018xhw}. This gives six roots two of them are real and the others are imaginary. These real roots are lengthy however, their behavior are drawn in Fig. \ref{Fig:4} \subref{fig:Hor}.  It is easy to check that  the degenerate horizon for the metric potential $B(r)$ given by Eq. (\ref{mpabds1})  is happened for a specific values for $(b,M,r)\equiv(0.242,1,0.333717999)$, respectively which corresponds to the Nariai BH. The degenerate behavior is shown is Fig. \ref{Fig:4} \subref{fig:metrd}. Figure \ref{Fig:4} \subref{fig:Hor} shows that  the horizon $r_-$ increasing with $M$ while $r_+$ decreasing.
Using Eq. (\ref{temp}) we draw the behavior of the Hawking temperatures  in Fig. \ref{Fig:5} \subref{fig:Temp} which shows that $T_{-}>T_{+}$. As Fig. \ref{Fig:5} \subref{fig:Temp} shows that the $T_-$ has an increasing positive temperature while $T_+$ has decreasing negative temperature.  The
Fig. \ref{Fig:5} \subref{fig:Temp} shows that $T_-=0$  at   $r_+ = r_d$. At $r_+ < r_d$, $T_-<0$ and  an ultracold BH is formed.

\begin{figure}
\centering
\subfigure[~Quasi local energy of BH (\ref{mpabds1})]{\label{fig:Enr}\includegraphics[scale=0.4]{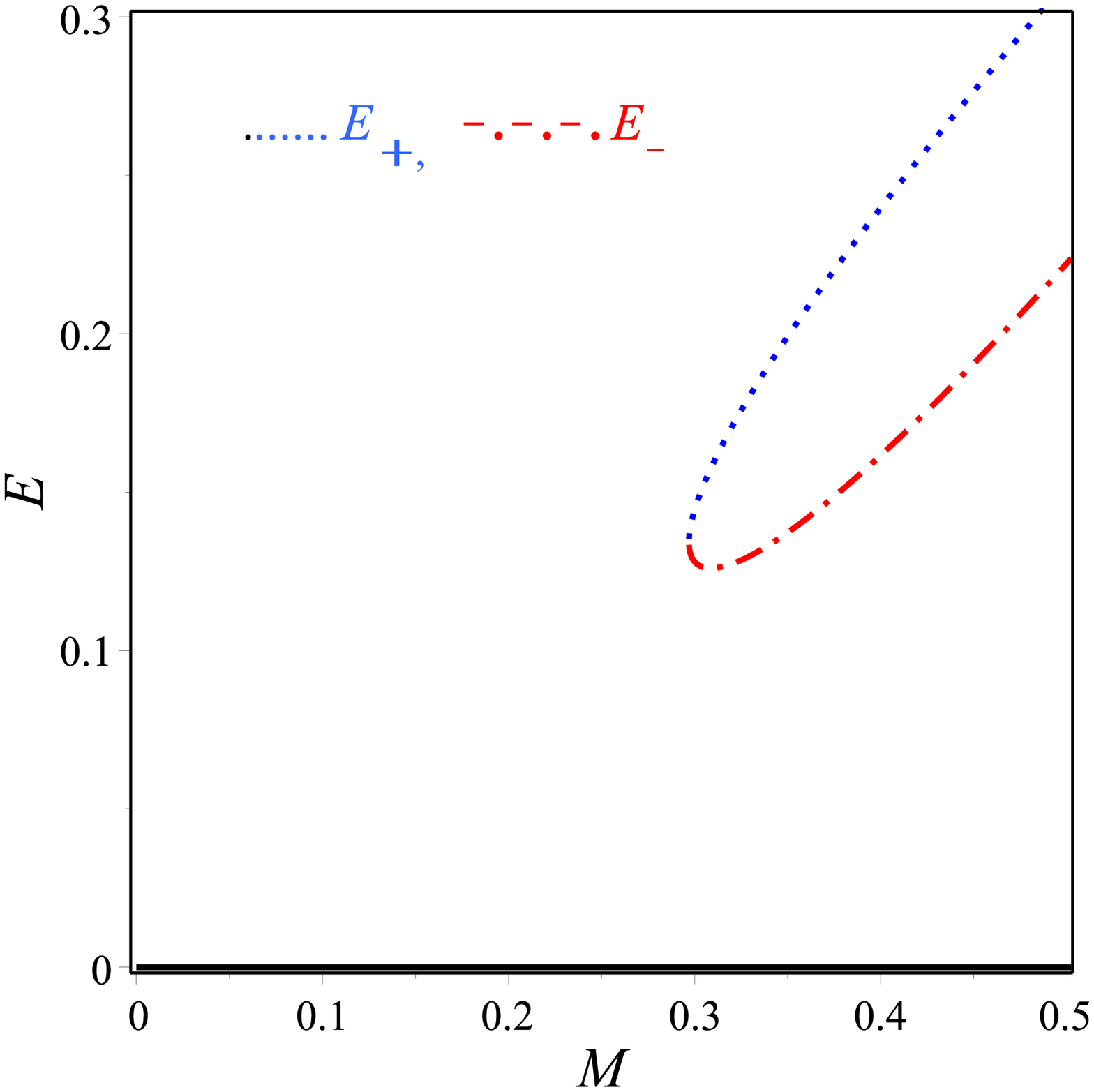}}
\subfigure[~Gibbs energy of BH (\ref{mpabds1})]{\label{fig:gib}\includegraphics[scale=0.4]{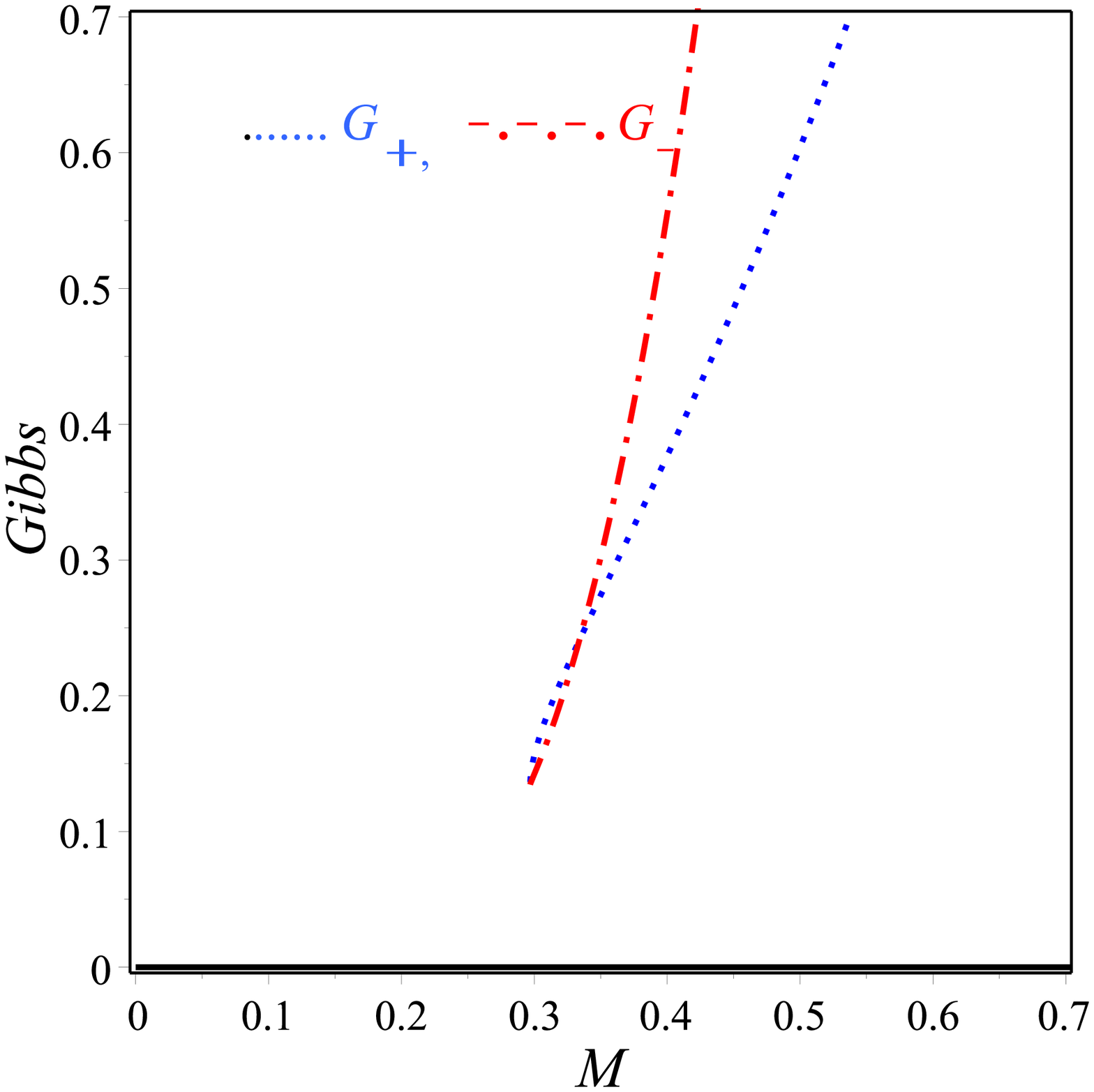}}
\caption[figtopcap]{\small{{Plot of the horizons given by Eq. (\ref{r1}) using $b=0.1$ which is consistent with the constrains (\ref{cons}).}}}
\label{Fig:6}
\end{figure}

Using Eq. (\ref{ent}) we draw the entropy  in Fig. \ref{Fig:5} \subref{fig:ent} showing  an increasing value for $\delta_-$ and decreasing value for $\delta_+$. Using Eq. (\ref{en}) we  draw the quasi-local energy  in Fig. \ref{Fig:6} \subref{fig:Enr} and show that it has a  positive increasing value for $E_\pm$. Finally, we use Eqs. (\ref{T1}),  (\ref{S1}) and  (\ref{E1}) in Eq.  (\ref{enr}) to calculate the Gibbs free energies. The behavior of these free energies are shown in Fig. \ref{Fig:6} \subref{fig:gib} which shows positive increasing values for $G_\pm$.

\subsection{Thermodynamics of the BH (\ref{mpab1}) that has  asymptote flat dS}
\begin{figure}
\centering
\subfigure[~The metric potential of BH (\ref{mpabds2})]{\label{fig:met}\includegraphics[scale=0.3]{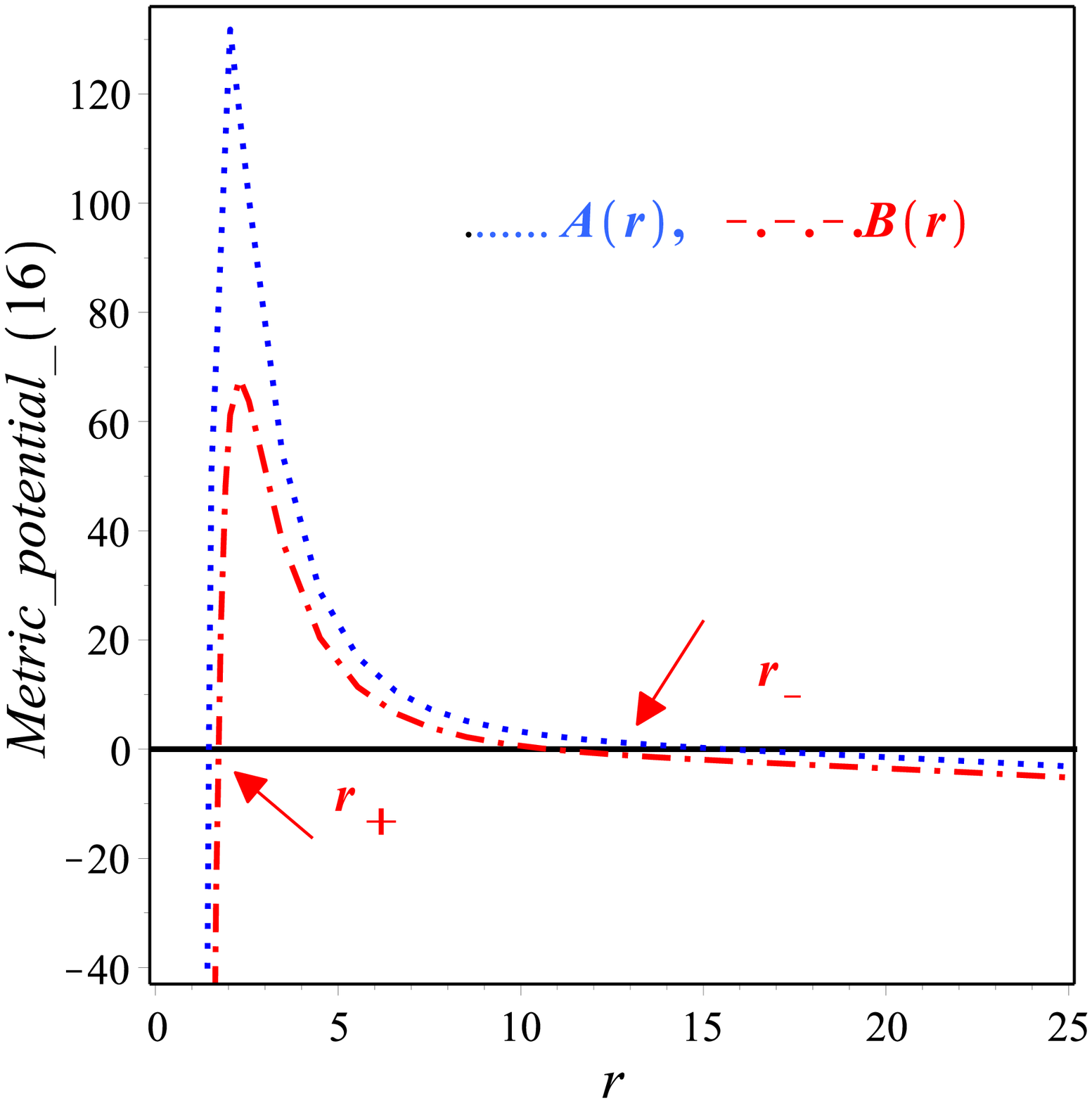}}
\subfigure[~The metric potential of BH (\ref{mpabds2})]{\label{fig:metrd}\includegraphics[scale=0.3]{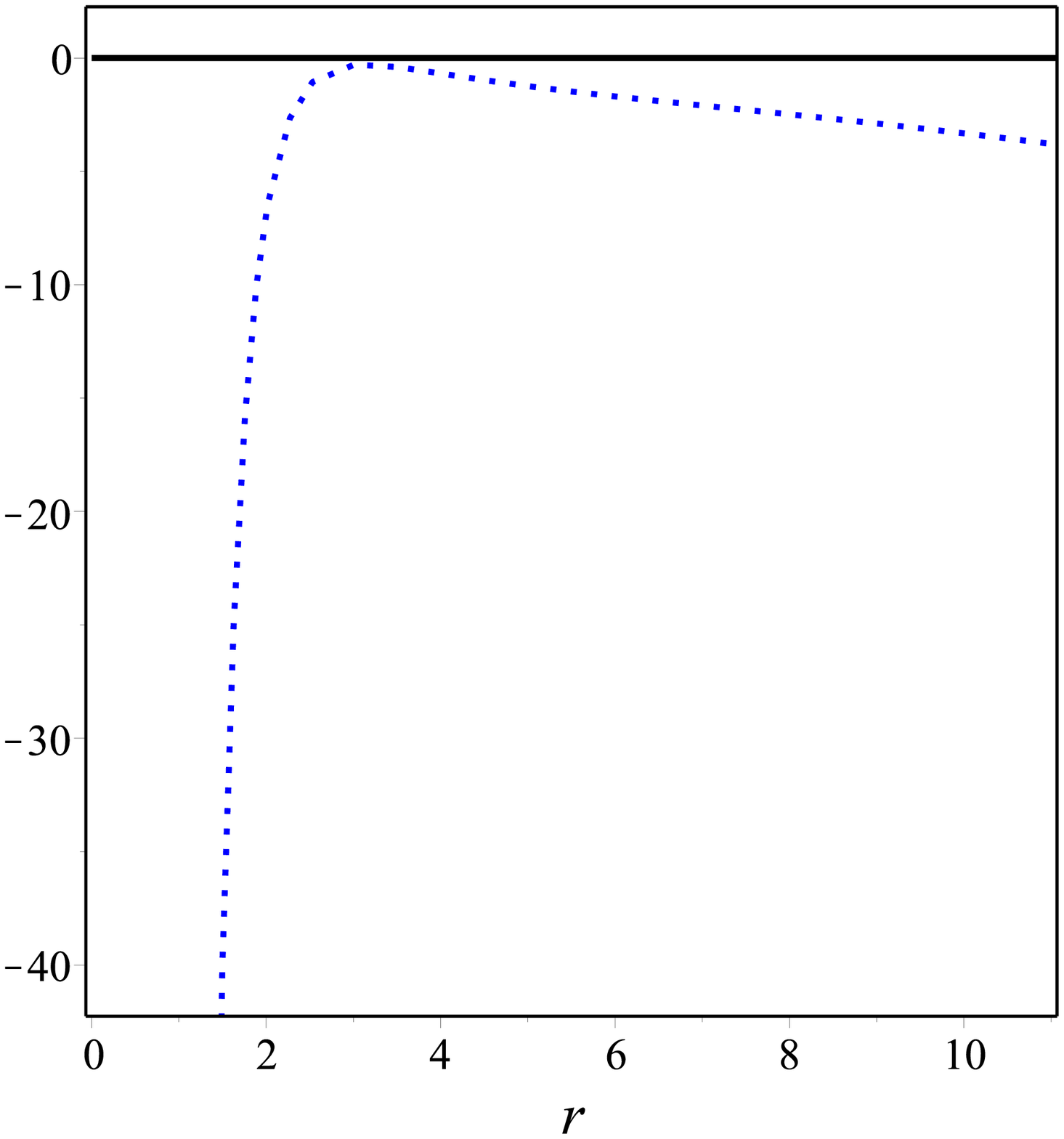}}
\subfigure[~Hawking temperature of BH (\ref{mpabds2})]{\label{fig:Hor}\includegraphics[scale=0.3]{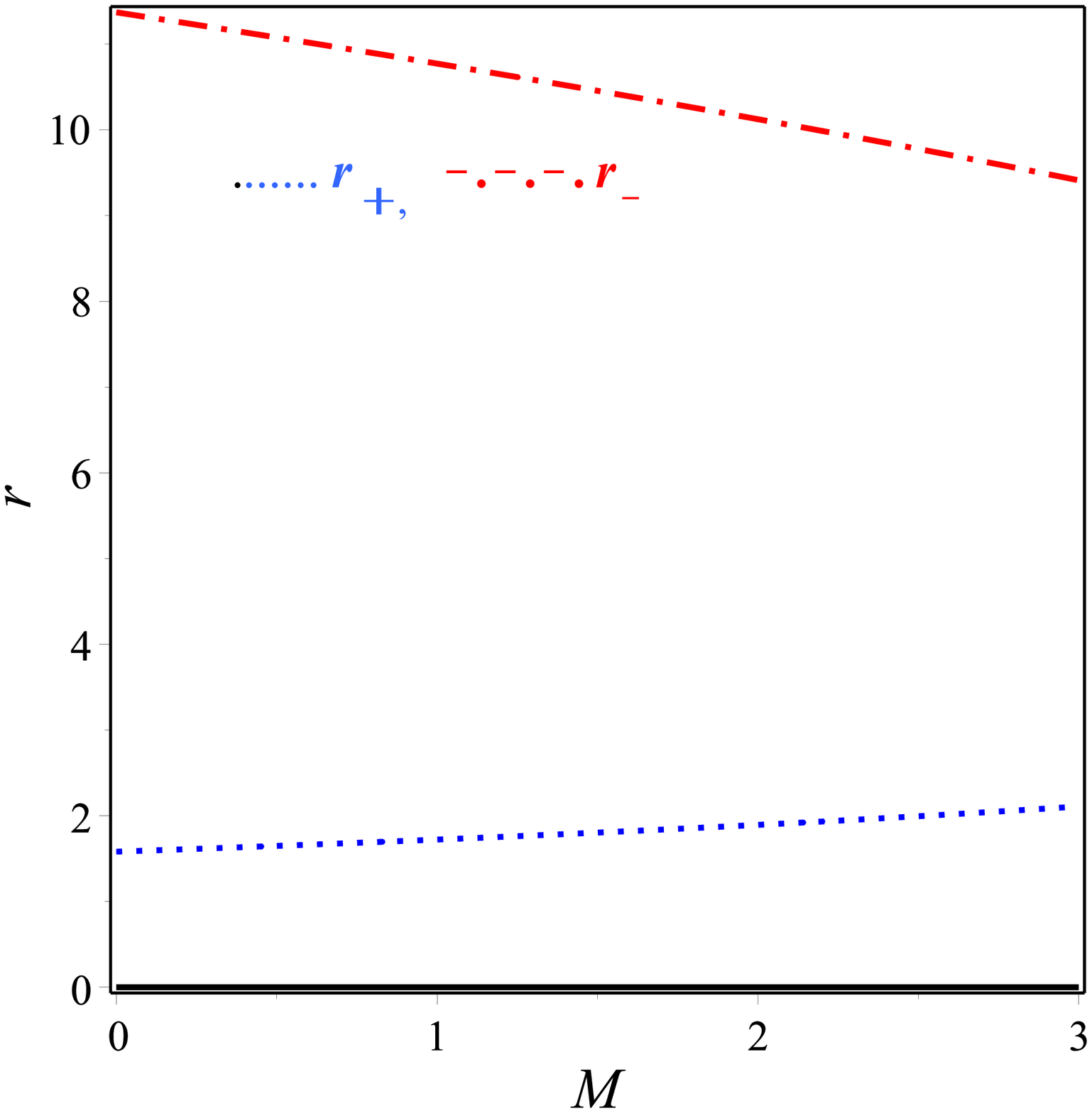}}
\caption[figtopcap]{\small{{Plot of the horizons given by Eq. (\ref{r1}) using $b=0.1$}}}
\label{Fig:7}
\end{figure}

\begin{figure}
\centering
\subfigure[~Hawking temperature of BH  (\ref{mpabds1})]{\label{fig:Temp}\includegraphics[scale=0.4]{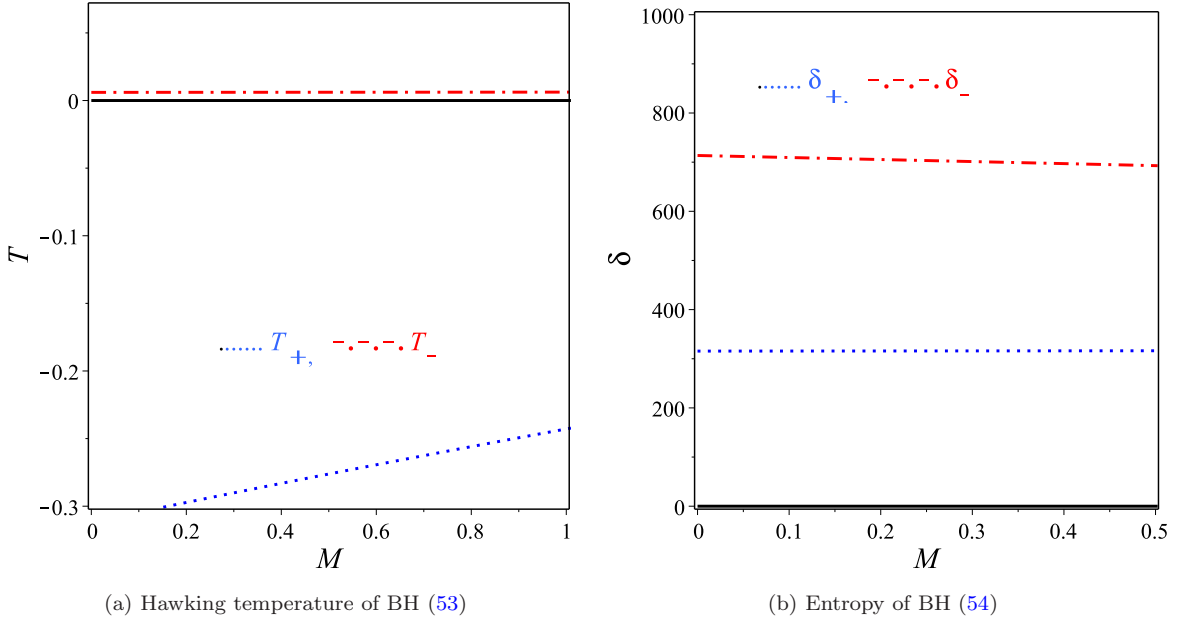}}
\subfigure[~Entropy of BH (\ref{mpabds2})]{\label{fig:ent}\includegraphics[scale=0.4]{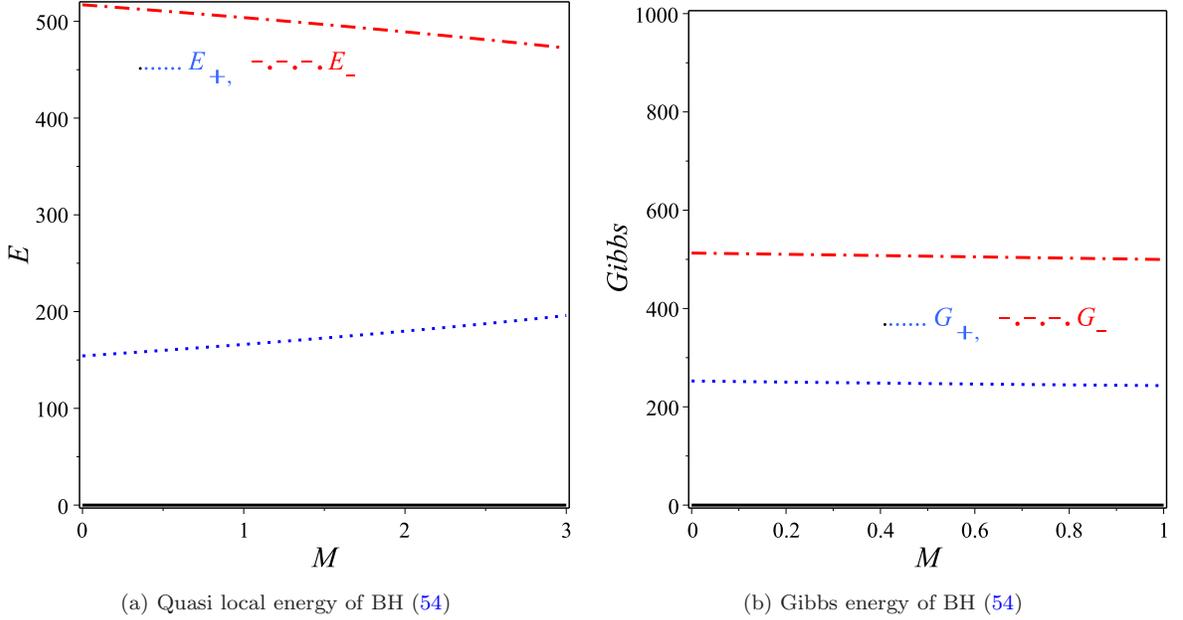}}
\caption[figtopcap]{\small{{Plot of the horizons given by Eq. (\ref{r1}) using $b=0.1$ which is consistent with the constrains (\ref{cons}).}}}
\label{Fig:8}
\end{figure}

The BH (\ref{mpabds2})  which is characterized  by the mass of the BH $M$ and the parameter $b$ and a cosmological effective constant\footnote{The metric potential when $\Lambda_\mathrm{eff.}=\frac{1}{3b^2}$ take the form
\begin{eqnarray}\label{mpabds2}
&& A(r)\approx -\Lambda_\mathrm{eff} r^2+1-\frac{2M}{r}-\frac{b^2}{r^2}+\frac{8b^4}{5r^3}+\frac{6b^2M}{r^3}-\frac{52b^4}{21r^4}+\cdots\,,\nonumber\\
 &&
B(r)\approx  -\Lambda_\mathrm{eff}r^2-1-\frac{2M}{r}-\frac{b^2}{r^2}+\frac{8b^4}{5r^3}-\frac{6b^2M}{r^3}-\frac{52b^4}{21r^4}+\cdots\,.\end{eqnarray}}. The metric potentials of the BH (\ref{mpabds2}) are figured  in Fig. \ref{Fig:7} \subref{fig:met}. From Fig. \ref{Fig:7} \subref{fig:met} we can easily see the  two horizons of the metric potentials $A(r)$ and  $B(r)$. To find the horizons of this BH, (\ref{mpabds2}), we put $A(r)=0$ in Eq. (\ref{mpabds2}) \cite{Wang:2018xhw}. This gives six roots two of them are real and the others are imaginary. These real roots are lengthy however, their behavior are drawn in Fig. \ref{Fig:7} \subref{fig:Hor}.  It is easy to check that  the degenerate horizon for the metric potential $B(r)$ given by Eq. (\ref{mpabds2})  is happened for a specific value for $(b,M,r)\equiv(0.242,1,0.333717999)$, respectively which correspond to Nariai BH. The degenerate behavior is shown is Fig. \ref{Fig:7} \subref{fig:metrd}. The Fig. \ref{Fig:7} \subref{fig:Hor} shows that  the horizon $r_-$ increasing with $M$ while $r_+$ decreasing.
Using Eq. (\ref{temp}) we draw the behavior of the Hawking temperatures  in Fig. \ref{Fig:8} \subref{fig:Temp} which shows that $T_{+}<T_{-}$. As Fig. \ref{Fig:8} \subref{fig:Temp} shows that the $T_-$ has an increasing positive temperature while $T_+$ has decreasing negative temperature.

\begin{figure}
\centering
\subfigure[~Quasi local energy of BH (\ref{mpabds2})]{\label{fig:Enr}\includegraphics[scale=0.4]{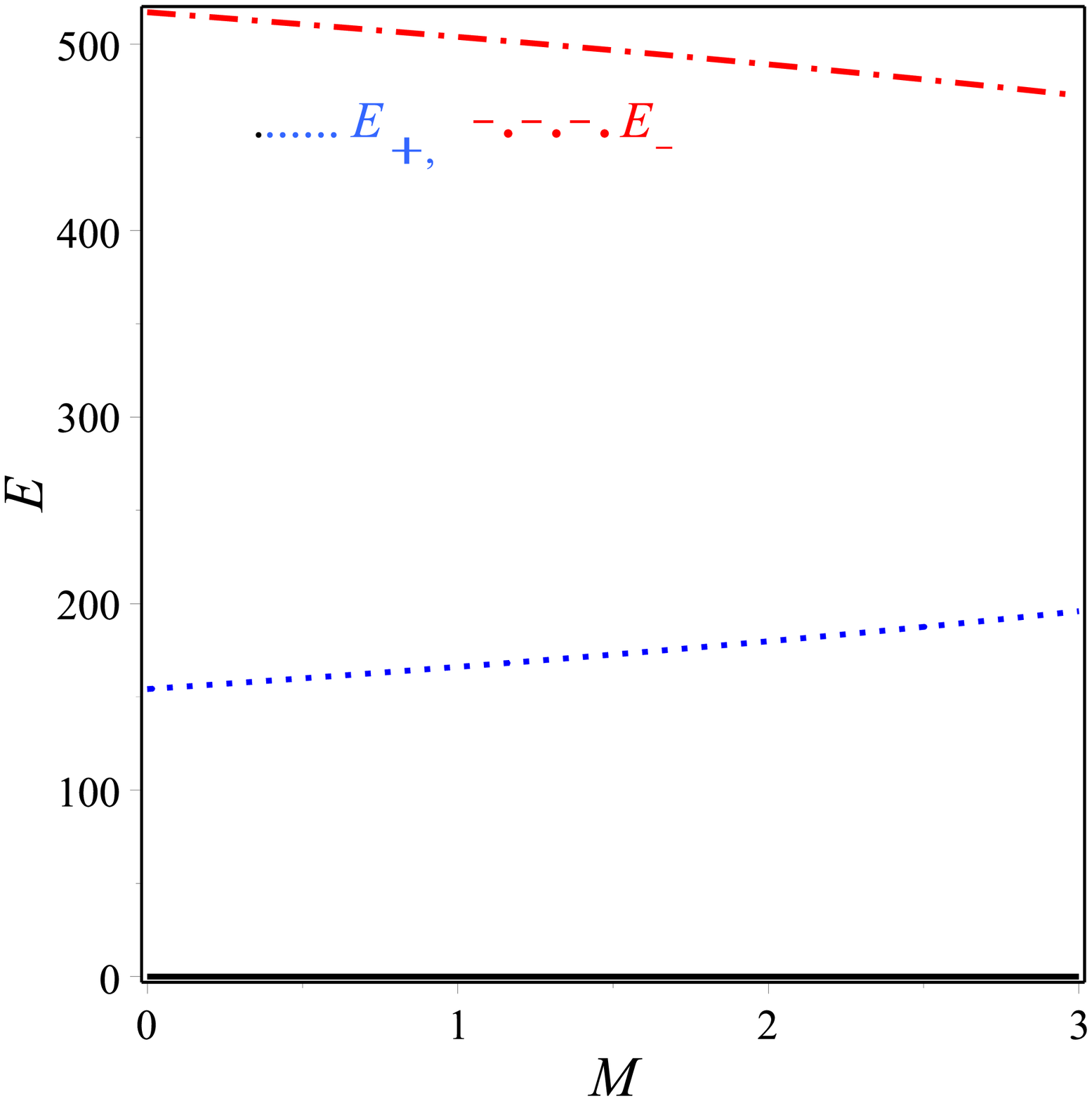}}
\subfigure[~Gibbs energy of BH (\ref{mpabds2})]{\label{fig:gib}\includegraphics[scale=0.4]{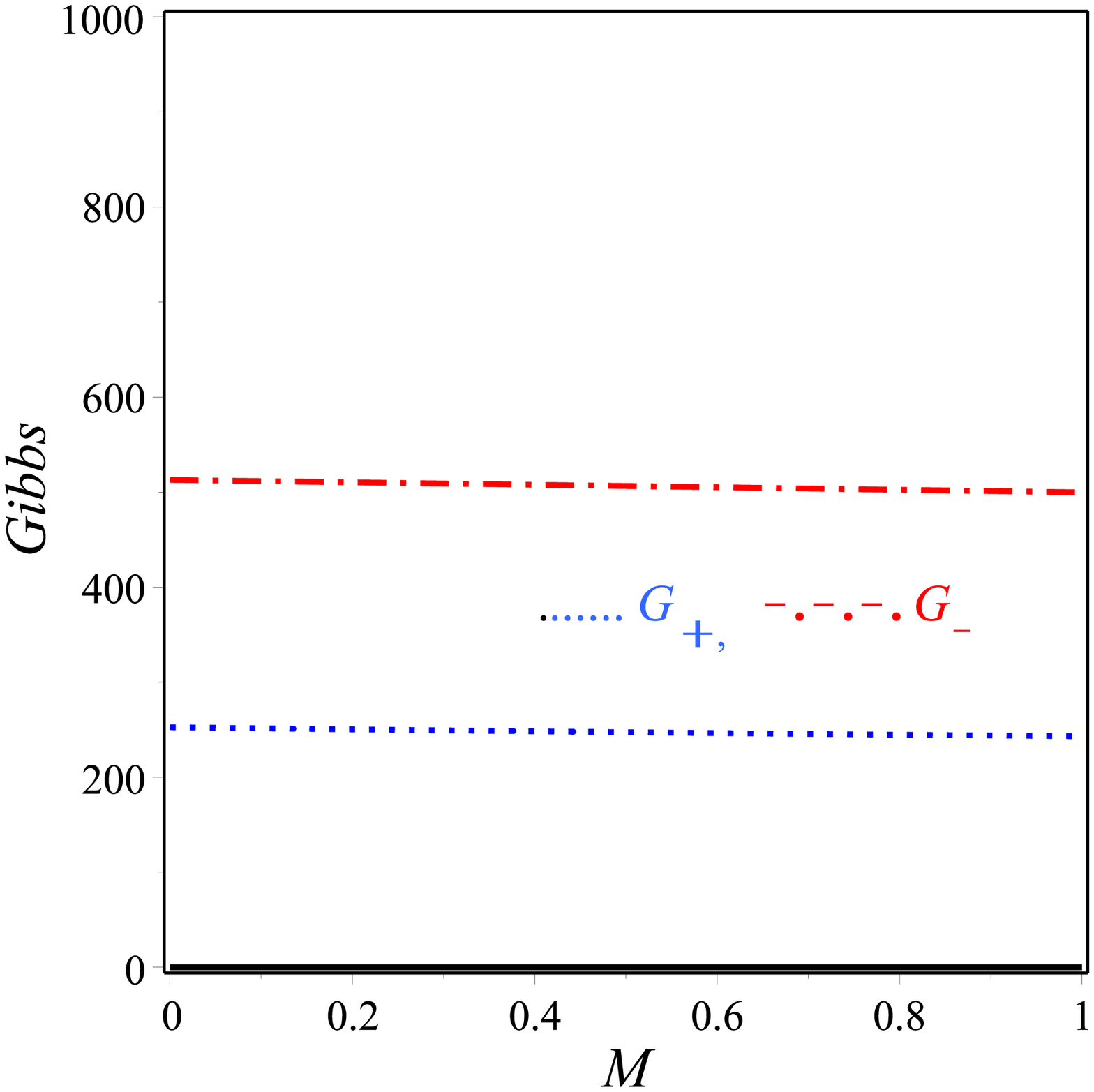}}
\caption[figtopcap]{\small{{Plot of the horizons given by Eq. (\ref{r1}) using $b=0.1$ which is consistent with the constrains (\ref{cons}).}}}
\label{Fig:9}
\end{figure}

Using Eq. (\ref{ent}) we draw the entropy  in Fig. \ref{Fig:8} \subref{fig:ent} showing that it has a positive value for $\delta_\pm$. From Eq. (\ref{en}) we calculate and draw the quasi-local energy in Fig. \ref{Fig:9} \subref{fig:Enr}  showing that it has a  positive increasing value for $E_\pm$. Finally, we use Eqs. (\ref{T1}),  (\ref{S1}) and  (\ref{E1}) in Eq.  (\ref{enr}) to calculate the Gibbs free energies. The behavior of these free energies are shown in Fig. \ref{Fig:9} \subref{fig:gib} which shows positive increasing values for $G_\pm$.
\subsection{First law of thermodynamics of the BHs  (\ref{mpab}) and (\ref{mpab1})}
It is worth mentioning to examine the verification  of  the first law for the BHs  (\ref{mpab}) and (\ref{mpab1}). Accordingly, we will apply this law on $f(R)$ using the following form \cite{Zheng_2018}
\begin{equation}
dE=Td\delta-PdV,\label{1st}
\end{equation}
where $E$ is the quasi-local energy, $\delta$ is the Bekenstein-Hawking entropy, $T$ is the Hawking temperature, $P$ is the radial component of the stress-energy tensor that serves as a thermodynamic pressure $P=T_r{}^r\mid_{\pm}$ and $V$ is the geometric volume. In the frame of $f(R)$ gravitational theory the pressure is defined as \cite{Zheng_2018}
\begin{equation}
P=-\frac{1}{8\pi}\left\{\frac{F}{r_{\pm}{}^2}+\frac{1}{2}(f-RF)\right\}+\frac{1}{4}\left(\frac{2F}{r_{\pm}}+F'\right)T\,.\label{1st}
\end{equation}
For the flat spacetime (\ref{mpab}) if we neglect $O\Big(\frac{1}{r^4}\Big)$ to make the calculations more applicable  for we get\footnote{When we neglect the terms of order $O\Big(\frac{1}{r^4}\Big)$ and when $A(r)=0$ we get three roots one of them only has positive value while the other two are imaginary.}

 \begin{eqnarray}
 &&r=\frac{2M}{3}+\frac{\sqrt[3]{1000m^3-2700b^4-9000mb^2+75\mho}}{15}+\frac{5(3b^2+4m^2)}{3\sqrt[3]{1000m^3-2700b^4-9000mb^2+75\mho}}\,,\qquad E=\frac{r^4-b^2r^2+4b^4}{2\,r^3}\,,\nonumber\\
&& \delta=\pi(r^2+2b^2), \qquad T=\frac{5mr^2+5b^2r-45b^2m-12b^4}{10\,\pi\, r^4}, \qquad P=\frac{10mr^2-5r^3+5b^2r-90mb^2-24b^4}{40\, \pi\, r^5}\,,
\end{eqnarray}
where $\mho=\sqrt{14100m^2b^4-75b^6-3600m^4b^2+8640mb^6+1296b^8-960m^3b^4}$. Substituting the form of $r$ into the thermodynamical quantities we get
 \begin{eqnarray} \label{1stv}
&& dE=\frac{4m^4+b^2m^2+b^4}{8\,m^4}\,,\qquad  \delta=\frac{\pi(40m^4-20m^2b^2-8mb^4-25b^4}{10\,m^3}, \qquad T=\frac{4mb^4-20m^4+35b^4+15m^2b^2}{160\,\pi\, m^5}\,,\nonumber\\
 && P=-\frac{b^2(30m^2+75b^2+8mb^2)}{640\, \pi \,m^6}\,.
\end{eqnarray}
If we use Eq. (\ref{1stv}) in (\ref{1st}) We can verify the first law of thermodynamics  for the BH (\ref{mpab}).

If one  repeats the same procedure for the BH (\ref{mpab1}) one can verify the first law of thermodynamics provided that we neglect all the quantities containing $b$ to make the calculation easier to carry out.
\section{The stability of  the  BHs  in $f(R)$ gravity}\label{S616}
 To study the stability of the above solutions, we recast  $\mathit{f(R)}$ gravity using scalar-tensor theory.
The action given by Eq. (\ref{a2}) can be recast as \cite{DeFelice:2011ka}
\begin{equation}
{\mathop{\mathcal{ I}}}=\frac{1}{2\kappa}\int d^{4}x\sqrt{-g}\,[\psi\, \mathbf{R}-V(\psi)],\label{action2}
\end{equation}
where $\psi$ is a scalar field   coupled to the Ricci scalar $\mathbf{(R)}$ and $V(\psi)$ is the potential (see \cite{Capozziello:2011et,DeFelice:2011ka} for details).
To discuss the perturbation we use  the spherically symmetric metric as
\begin{equation}
ds^{2}=\mathit{g_{\mu\nu}^{0}}dx^{\mu}dx^{\nu}=-A(r)\, dt^{2}+\frac{dr^{2}}{B(r)}+ r^{2} ( d\theta^{2} +\sin^2\theta\, d\phi^{2}).
\end{equation}
with $\mathit{g_{\mu\nu}^{0}}$ being the background metric.
We are going to check BH solutions, (\ref{mpab}) and (\ref{mpab1}), using  the linear  perturbations,   if their
background metrics are stable or not.  Moreover, we are going to investigate  the value of the speed of propagation. For this theory, the background
equation reads
\begin{equation}
  V =- {\frac{4B\, \psi'}{r}}-{\frac{2\psi\, B A'}{Ar}}-\frac{\psi'B A'}{A}+{\frac{2\psi}{{r}^{2}}}-{\frac{2B\, \psi}{{r}^{2}}}\, , \qquad \ \qquad \psi'' =-\frac{\psi'B'}{2B}-\frac{\psi B'}{r B}+\frac{\psi'A'}{2A}+\frac{\psi A'}{rA}\,,\qquad \qquad   \mathbf{R} =\frac{dV}{d\psi}\,,\label{eq:Ueq}\\
\end{equation}
 where $'$  means the differentiation  w.r.t. the radial coordinate, $r$.

\subsection{Brief description of Regge-Wheeler-Zerilli construction}

Following the workings by Regge,  Wheeler \cite{PhysRev.108.1063} and Zerilli \cite{PhysRevLett.24.737} to decompose
the metric perturbations subject to the  transformation properties
of 2-dimensional rotations.
Due to the fact that Regge, Wheeler and Zerilli studied
the perturbations of   the Schwarzschild BH  in GR we can apply it
 to  the BHs of  $\mathit{f(R)}$.

We  will assume the perturbation of $\mathit{g_{\mu\nu}}$ as
\[\mathit{g_{\mu\nu}}=\mathit{ g_{\mu\nu}^{\mathrm{0}}}+\mathit{h_{\mu\nu}},\]  with $\mathit{ h_{\mu\nu}}$ being  small quantities. In the lower order  we assume the perturbations to be small i.e., $\mathit{g_{\mu\nu}^{\mathrm{0}}}>>\mathit{h_{\mu\nu}}$. Therefore, on  two-dimensional
rotations, $\mathit{h_{tt}}, \mathit{h_{tr}}$ and $\mathit{h_{rr}}$ transform as  $\mathit{h_{ta}}$
and $\mathit{h_{ra}}$ transforms as a vector and $\mathit{h_{ij}}$ transforms as a tensor (where $i$ and $j$ are either $\theta$ or $\phi$). It is will known that  $\Phi$  can be expressed  as:
 \begin{equation}
\Theta(t,r,\theta,\phi)=\sum_{\ell,m}\Theta_{\ell m}(t,r)Y_{\ell m}(\theta,\varphi).\label{scalar-decomposition}
\end{equation}
Thus, the apparent solution becomes  independent of the index  $m$ and  takes the form:
\begin{equation} \Delta_{\theta, \phi}Y_{\ell}(\theta, \phi)=-\ell(\ell+1)Y_{\ell}(\theta, \phi).
\end{equation}
 As for the vector $C_{a}$ one can decompose it into two parts, a divergence part
and a non-divergence one as:
 \begin{equation}
C_{a}(t,r,\theta,\phi)=\nabla_{a}\Theta_{1}+E_{a}^b\nabla_b\Theta_{2},
\end{equation}
such that  $\Theta_{1}$ and $\Theta_{2}$ are two scalars and $E_{ab}\equiv\sqrt{\det\Omega}~\epsilon_{ab}$
where $\Omega_{ab}$ is the two-dimensional metric
and $\epsilon_{ab}$ is  completely skew-symmetric with
$\epsilon_{\theta \varphi}=1$. In this study,  $\nabla_{a}$ is the covariant derivative w.r.t. the metric $\Omega_{ab}$.
Since $C_{a}$  is  two-component vector and
can be described by  $\Theta_{1}$ and $\Theta_{2}$. Thus
 the scalar decomposition (\ref{scalar-decomposition}) can be applied  to $\Theta_{1}$
and $\Theta_{2}$ to decompose  $C_a$.

For  $S_{ab}$ which is symmetric one can decompose it as
 \begin{equation}
S_{ab}(t,r,\theta,\phi)=\nabla_{a}\nabla_{b}\Theta_{1}+\gamma_{ab}\Theta_{2}+\frac{1}{2}\left(E_{a}{}^{c}\nabla_{c}\nabla_{b}\Theta_{3}+
E_{b}{}^{c}\nabla_{c}\nabla_{a}\Theta_{3}\right),
\end{equation}
with $\Theta_{1},~\Theta_{2}$ and $\Theta_{3}$ are scalars. Since $S_{ab}$ have three
independent components that completely
describe $S_{ab}$. Therefore, one can  use the scalar decomposition
(\ref{scalar-decomposition}) to $\Theta_{1},~\Theta_{2}$ and $\Theta_{3}$
to decompose  $S_{ab}$. We mention  $O_{ab}$ by
 the odd-type variables and the rest by even-type
ones. The use of these methods  has an advantage that in the linearized form  the
odd-type and even-type are separated which makes us study them separately.

\subsection{Perturbations of $f(R)$ gravity using the odd-modes}

Using the Regge-Wheeler formalism, the odd-type metric perturbations
can have the form
 \begin{eqnarray}
 &  & \mathit{h_{tt}}=0,~~~\mathit{h_{tr}}=0,~~~\mathit{h_{rr}}=0,\\
 &  & \mathit{h_{ta}}=\mathit{\sum_{\ell, m}h_{\mathrm{0},\ell m}(t,r)E_{ab}\partial^{b}Y_{\ell m}(\theta,\varphi)},\\
 &  & \mathit{h_{ra}=\sum_{\ell, m}h_{\mathrm{1},\ell m}(t,r)E_{ab}\partial^{b}Y_{\ell m}(\theta,\varphi)},\\
 &  & \mathit{h_{ab}=\frac{1}{2}\sum_{\ell, m}h_{\mathrm{2},\ell m}(t,r)\left[E_{a}^{~c}\nabla_{c}\nabla_{b}Y_{\ell m}(\theta,\varphi)+E_{b}^{~c}\nabla_{c}\nabla_{a}Y_{\ell m}(\theta,\varphi)\right]}.
\end{eqnarray}
Thus, implementing the gauge transformation $x^{\mu}\to x^{\mu}+\xi^{\mu}$  we can put  some of the metric perturbations equal zero due to the fact  that all them
 need not to be physical ones. We use the following transformation for the odd-type perturbation:
 \begin{equation}
\mathit{\xi_{t}}=\mathit{\xi_{r}}=0,~~~\mathit{\xi_{a}}=\mathit{\sum_{\ell m}\Lambda_{\ell m}(t,r)E_{a}^{~b}\nabla_{b}Y_{\ell m}},
\end{equation}
with $\Lambda_{\ell m}$  can put $h_{\mathrm{2},\ell m}$ equal zero
(the Regge-Wheeler gauge).  Using this  method, we can fix $\Lambda_{\ell m}$  and the action of odd modes takes the form \cite{PhysRev.108.1063}
\begin{eqnarray}
&&I_{\mathrm{ odd}}=\frac{1}{2\kappa} \sum_{\ell,m}\int dt\, dr\,{\cal I}_{{ odd}}=\frac{1}{4\kappa} \sum_{\ell,m}\int dt\, dr\,j^{2}\bigg[\psi\sqrt{\frac{B}{A}}{\left({\dot{h}_{\mathrm{1}}}-h_\mathrm{0}'\right)}^{2}+\frac{4h_\mathrm{0}{\dot{h}_\mathrm{1}} \psi}{r}\sqrt{\frac{B}{A}}+ \frac{h_\mathrm{0}^{2}}{r^{2}}\Big[2\Bigg(r\Bigg\{\psi \sqrt{\frac{B}{A}}\Bigg\}' +\psi\sqrt{\frac{B}{A}}\Bigg)\nonumber\\
&&+\frac{(j^2-2)\psi}{\sqrt{AB}} \Big]-\frac{\,(j^{2}-2)\,\sqrt{AB}\, \psi\, h_\mathrm{1}^{2}}{r^{2}}\bigg],\label{odd-action}
\end{eqnarray}
with $j^2=l(l+1)$. Equation (\ref{odd-action}) is  independent of $\dot{h}_0$  and therefore, the variation of it w.r.t. $h_{0}$ gives
 \begin{equation}
\Big[\psi(h_\mathrm{0}'-\dot{h}_\mathrm{1})\sqrt{\frac{B}{A}}\Big]'= \frac{1}{r^{2}}\Bigg[2\Bigg(r\Bigg\{\psi \sqrt{\frac{B}{A}}\Bigg\}' +\psi\sqrt{\frac{B}{A}}\Bigg)+\frac{(j^2-2)\psi}{\sqrt{AB}} \Bigg]\, h_\mathrm{0}+\frac{2 \sqrt{\frac{B}{A}}\psi\,{\dot{h}_\mathrm{1}}}{r}\,.\label{cons}
\end{equation}
Equation (\ref{cons}) has no solution for $h_{0}$ therefore, (\ref{odd-action}), turns to be
\begin{eqnarray}
&&{I}_{\mathrm{ odd}}=\frac{ j^{2}\, \psi\sqrt{\frac{B}{A}}}{2}{\left({\dot{h}_\mathrm{1}}-h_\mathrm{0}'+\frac{2\,{h_\mathrm{0}}}{r}\right)}^{2}-\frac{j^{2}\Big(\psi\sqrt{\frac{B}{A}}
+r\Big[\sqrt{\frac{B}{A}}\psi\Big]'\Big)
\,h_\mathrm{0}{}^2}{r^2}+\frac{j^{2}\,h_\mathrm{0}^{2}}{2r^{2}}\left[2\Bigg(r\Bigg\{\psi \sqrt{\frac{B}{A}}\Bigg\}' +\psi\sqrt{\frac{B}{A}}\Bigg)+\frac{(j^2-2)\psi}{\sqrt{AB}} \right]\nonumber\\
&&-\frac{j^{2}\,(j^{2}-2)\,\sqrt{AB}\, \psi\, h_{1}^{2}}{2r^{2}}\,.\label{eq:Lodd2}
\end{eqnarray}
Using the Lagrange multiplier, $q$,  Eq.~(\ref{eq:Lodd2}) can be rewritten as follows
\begin{eqnarray}
&&{I}_{{  odd}}=\frac{j^{2}\, \psi\sqrt{\frac{B}{A}}}{2}\left[2\, q\left(\dot{h}_{\mathrm{1}}-h'_{\mathrm{0}}+{\frac{2\,h_{{0}}}{r}}\right)-q^{2}\right]-\frac{j^{2}\Big(\psi\sqrt{\frac{B}{A}}
+r\Big[\sqrt{\frac{B}{A}}\psi\Big]'\Big)
\,h_\mathrm{0}{}^2}{r^2}+\frac{j^{2}\,h_\mathrm{0}^{2}}{2r^{2}}\Bigg[2\Bigg(r\Bigg\{\psi \sqrt{\frac{B}{A}}\Bigg\}' +\psi\sqrt{\frac{B}{A}}\Bigg)\nonumber\\
&&+\frac{(j^2-2)\psi}{\sqrt{AB}} \Bigg]-\frac{j^{2}\,(j^{2}-2)\,\sqrt{AB}\, \psi\, h_{1}^{2}}{2r^{2}}\,.\label{eq:Lodd3}
\end{eqnarray}
Equation  (\ref{eq:Lodd3}) can be written as
\begin{eqnarray}
h_\mathrm{1} &=& -\frac{r^2\,\dot{q}}{(j^{2}-2)A}\,,\label{eq:oddh1}\\
h_\mathrm{0}&=&  \frac{r\Bigg[\Big\{\psi\sqrt{\frac{B}{A}}+2r\Big[\sqrt{\frac{B}{A}}\psi\Big]'\Big\}q+2q'r\psi\sqrt{\frac{B}{A}}\Bigg]}{2j^{2}  \Bigg[\psi\sqrt{\frac{B}{A}}+r\Big[\sqrt{\frac{B}{A}}\psi\Big]'-\Bigg(r\Bigg\{\psi \sqrt{\frac{B}{A}}\Bigg\}' +\psi\sqrt{\frac{B}{A}}+\frac{(j^2-2)\psi}{2\sqrt{AB}}\Bigg)\Bigg]}\:.\label{eq:oddh0}
\end{eqnarray}
 Equation (\ref{eq:oddh1}) relates the physical modes, $h_{0}$ and $h_{1}$,  to $q$. Therefore, when the auxiliary field $q$ is known the  quantities  $h_\mathrm{0}$ and $h_\mathrm{1}$
are also known.  Using the physical quantities $h_\mathrm{0}$ and $h_\mathrm{1}$ into the Eq. (\ref{odd-action})  and by carrying out
the integration by parts to the term proportional to $q'\, q$, we get
\begin{equation}
{I}_{\mathrm{odd}}=\frac{j^2r^2\psi\sqrt{\frac{B}{A^3}}}{2(j^2-2)}\,\dot{q}^{2}-\frac{j^2B\, \psi^2 \,q'^{2}}{4A\Bigg[\Bigg(r\Bigg\{\psi \sqrt{\frac{B}{A}}\Bigg\}' +\psi\sqrt{\frac{B}{A}}+\frac{(j^2-2)\psi}{2\sqrt{AB}}\Bigg)-\psi\sqrt{\frac{B}{A}}+r\Big[\sqrt{\frac{B}{A}}\psi\Big]'\Bigg]}-\alpha{}^{2}\, q^{2}\,,\label{eq:LoddF}
\end{equation}
 where
\begin{eqnarray}
\alpha^{2}&=&\frac{a_1r^2\Big[r^2a'_1a'_3-r^2a''_1a_3+2a_1a_3+4a'_1{}^2+r^2a_3{}^2-2a_1a''_1+2ra_1a'_3-4rs'_1a_3\Big]}{(2a_1+2ra'_1-r^2a_3)^2}\,,
\end{eqnarray}
and
\begin{equation}
a_1=\frac{j^2\psi \sqrt{B}}{2\sqrt{A}},\qquad a_2=\frac{2j^2\psi(j^2-2)\sqrt{AB}}{r^2}, \qquad a_3=\frac{j^2}{r^2}\Bigg(\frac{\psi \sqrt{B}}{\sqrt{A}}+r\Bigg\{\frac{\psi \sqrt{B}}{\sqrt{A}}\Bigg\}'+\frac{(j^2-2)\psi}{2\sqrt{AB}}\Bigg)\,.\nonumber\\
\end{equation}
Equation (\ref{eq:LoddF}),  shows  that it is free from ghosts and gives
\[
j^2\geq2\,,\qquad{  \mbox{and} \, \,\qquad}\psi \sqrt{\frac{B}{A^3}}\geq0\,.
\]
Therefore, for solutions that proportional to $e^{i(\omega t-kr)}$ where
$k$ and $\omega$ are large,  the radial dispersion relation is regarded as
\[
\omega^{2}=A\, B\, k^{2}\,.
\]
  Thus
the  radial speed reads

\[
c_{\mathrm{ odd}}^{2}=\left(\frac{dr_{*}}{d\tau}\right)^2=1\,,
\]
 where we used the radial tortoise coordinate ($dr_{*}^{2}=dr^{2}/B$)
and the proper time ($d\tau^{2}=A\, dt^{2}$).

\subsection{Geodesic deviation}\label{S66}
 The trajectories of a   particle in a gravitational background are prescribed   by
 \begin{equation}\label{ge}
\mathit{ d^\mathrm{2} x^\sigma \over d\tau^\mathrm{2}}+ \left\{^\sigma_{ \mu \nu} \right \}
 {d x^\mu \over d\tau}{d x^\nu \over d\tau}=0,
 \end{equation}
 where $\tau$ is an affine  parameter along the geodesic. Equation (\ref{ge}) is the geodesic equations and their  deviation take the form \cite{1992ier..book.....D}
  \begin{equation} \label{ged}
 \mathit{{d^\mathrm{2} \xi^\sigma \over d\tau^\mathrm{2}}+ \mathrm{2}\left\{^\sigma_{ \mu \nu} \right \}
 {d x^\mu \over d\tau}{d \xi^\nu \over ds}+
 \left\{^\sigma_{ \mu \nu} \right \}_{,\ \rho}
 {d x^\mu \over d\tau}{d x^\nu \over d\tau}\xi^\rho=0},
 \end{equation}
where $\xi^\rho$ is the deviation 4-vector. Using  Eqs. (\ref{ge}) and (\ref{ged})  into the line-element (\ref{met12}) we get \begin{equation} \label{gedi}
\mathit{{d^\mathrm{2} t \over d\tau^\mathrm{2}}=0, \qquad {1 \over \mathrm{2}} A'(r)\left({d t \over
d\tau}\right)^\mathrm{2}-r\left({d \phi \over d\tau}\right)^\mathrm{2}=0, \qquad {d^\mathrm{2}
\theta \over d\tau^\mathrm{2}}=0,\qquad {d^\mathrm{2} \phi \over d\tau^\mathrm{2}}=0,}\end{equation} and for
and \begin{eqnarray}\label{ged1} && \mathit{ {d^\mathrm{2} \xi^\mathrm{1} \over d\tau^\mathrm{2}}+B(r)A'(r) {dt \over d\tau}{d
\xi^\mathrm{0} \over d\tau}-\mathrm{2}r B(r) {d \phi \over d\tau}{d \xi^\mathrm{3} \over
d\tau}+\left[{\mathrm{1} \over \mathrm{2}}\left(A'(r)B'(r)+B(r) A''(r)
\right)\left({dt \over d\tau}\right)^\mathrm{2}-\left(B(r)+rB'(r)
\right) \left({d\phi \over d\tau}\right)^\mathrm{2} \right]\xi^\mathrm{1}=0}, \nonumber\\
&& \mathit{ {d^\mathrm{2} \xi^\mathrm{0} \over
d\tau^\mathrm{2}}+{A'(r) \over A(r)}{dt \over d\tau}{d \zeta^1 \over d\tau}=0,\qquad {d^\mathrm{2} \xi^\mathrm{2} \over d\tau^\mathrm{2}}+\left({d\phi \over d\tau}\right)^\mathrm{2}
\xi^\mathrm{2}=0, \qquad \qquad  {d^\mathrm{2} \xi^\mathrm{3} \over d\tau^\mathrm{2}}+{\mathrm{2} \over r}{d\phi \over d\tau} {d
\xi^\mathrm{1} \over d\tau}=0,} \end{eqnarray} where $A(r)$ and $B(r)$ are defined by the Eq.  (\ref{mpab}) or Eq.  (\ref{mpab1}). Equations (\ref{gedi})  and (\ref{ged1}) are the geodesic and geodesic devations. Using
the circular orbit  \begin{equation} \mathit{\theta={\pi \over \mathrm{2}}, \qquad
{d\theta \over d\tau}=0, \qquad {d r \over d\tau}=0,}
\end{equation}
we get
\begin{equation}
 \mathit{\left({d\phi \over d\tau}\right)^\mathrm{2}={A'(r)
\over r(\mathrm{2}A(r)-rA'(r))}, \qquad \left({dt \over
d\tau}\right)^\mathrm{2}={\mathrm{2} \over \mathrm{2}A(r)-rA'(r)}.} \end{equation}

Equations (\ref{ged1}) can be rewritten as
\begin{eqnarray} \label{ged2} &&  \mathit{{d^\mathrm{2} \xi^\mathrm{1} \over d\phi^\mathrm{2}}+B(r)A'(r) {dt \over
d\phi}{d \xi^\mathrm{0} \over d\phi}-\mathrm{2}r B(r) {d \xi^\mathrm{3} \over
d\phi} +\left[{\mathrm{1} \over \mathrm{2}}\left(A'(r)B'(r)+B(r) A''(r)
\right)\left({dt \over d\phi}\right)^\mathrm{2}-\left(B(r)+rB'(r)
\right)  \right]\zeta^\mathrm{1}=0,} \nonumber\\
&&\mathit{{d^\mathrm{2} \xi^\mathrm{2} \over d\phi^\mathrm{2}}+\xi^\mathrm{2}=\mathrm{0}, \qquad {d^\mathrm{2} \xi^\mathrm{0} \over d\phi^\mathrm{2}}+{A'(r) \over
A(r)}{dt \over d\phi}{d \xi^\mathrm{1} \over d\phi}=\mathrm{0},\qquad {d^\mathrm{2} \xi^\mathrm{3} \over d\phi^\mathrm{2}}+{\mathrm{2} \over r} {d \xi^\mathrm{1} \over
d\phi}=\mathrm{0}.} \end{eqnarray}
The second equation  of (\ref{ged2}) shows that it possesses  a simple harmonic motion which  means it has a  stable motion. We can assume the solution of the remaining of Eq. (\ref{ged2}) to be  \begin{equation} \label{ged3}
\mathit{\xi^\mathrm{0} = \zeta_\mathrm{1} e^{i \sigma \phi}, \qquad \xi^\mathrm{1}= \zeta_\mathrm{2}e^{i \sigma
\phi}, \qquad and \qquad \xi^\mathrm{3} = \zeta_\mathrm{3} e^{i \sigma \phi},}\end{equation}  where
$\zeta_1, \zeta_2$ and $\zeta_3$ are constants and    $\phi$  should be determined. Substituting (\ref{ged3}) in
(\ref{ged2}),  we get  \begin{equation} \label{con1}  \mathit{\displaystyle\frac{\mathrm{3}ABA'-\omega^\mathrm{2}A'A-\mathrm{2}rBA'^\mathrm{2}+rABA''}{AB'}>0,} \end{equation} which is the stability condition. Equation (\ref{con1}) for the BH (\ref{mpab})  can be rewritten as  \begin{equation}\label{stc}
\mathit{\mathrm{24}\,{M}^\mathrm{2}r{b}^\mathrm{2}+\mathrm{84}\,{b}^{\mathrm{2}}{M}^\mathrm{3}-\mathrm{29}\,M{r}^\mathrm{2}{b}^\mathrm{2}+M{r}^\mathrm{4}
+\mathrm{4}
\,{M}^\mathrm{2}{r}^\mathrm{3}-\mathrm{12}\,{M}^\mathrm{3}{r}^\mathrm{2}+\mathrm{2}\,{b}^\mathrm{2}{r}^\mathrm{3}>0,}\end{equation}
which is the  stability condition for the solution (\ref{mpab}) and when $b=0$ we get $r>2M$ which is the stability condition of the Schwarzschild spacetime \cite{Misner:1974qy}.

\section{ Discussion and conclusions }\label{S77}
Spherically symmetric spacetime is considered as an ingredient tool for BH physics due to the fact that its basic properties can be investigated easily  \cite{Chakraborty:2016lxo}. Previously there were many spherically symmetric BH solutions provided a specific form of $f(R)$  gravity and equal metric potential \cite{Nashed:2019tuk,Elizalde:2020icc}. In the present study, we use a spherically symmetric spacetime having unequal  metric potentials and without assuming any form of the metric potential.  Before we continue we emphasis on the following points:\hspace{0.2cm}\\
i) From the trace equation of the field equation of $f(R)$ we isolate the form of $f(R)$ in one side.\hspace{0.2cm}\\
ii) We use this form of $\mathit{f(R)}$ in the field equations (\ref{f3ss}), (\ref{f3}) and got a form of the field equations that contains only the first derivative of $\mathit{f(R)}$ w.r.t. $R$. Thus, we have applied the above mentioned form as on  Eq. (\ref{met12}) and derived the differential equations that governed such a system. We have succeeded to solve such a system analytically and derive the form of the metric potentials in addition to the form of $\mathit{f(R)}$.

Moreover, it has been revealed that the associated  BH-solution depends only on a convolution function that it is responsible to make it different from GR BH solution. Under some constraints, if this convolution function is equal to zero we discover the  Schwarzschild BH of the Einstein GR. Therefore, the effect of higher curvature of $\mathit{f(R)}$ gravity is restored on this convolution function. To understand the physics of this original solution we asymptote the convolution function up to certain order, fourth-order. The most beautiful thing in this asymptote is the fact that we get two constants one can be related to the cosmological constant and the other is the one responsible to make a deviation from GR BH.   So we classified our asymptote to two classes with the constant that is related to the cosmological constant and the other without this constant. As for the form of the metric potentials without the cosmological constant we restore all the higher-order correction to one constant and show that the metric potentials asymptote as a flat spacetime. Also, we show that if this constant is vanishing  we  discover the Schwarzschild BH of GR. As for the second asymptote that includes the cosmological constant the line element behaves asymptotically as AdS/dS spacetime. It is  important to compare our results with the ones obtained  by Jaime et al \cite{Jaime:2010kn}: Jaime et al
studied $\mathit{f(R)}$ models that fulfil  two conditions: $\frac{\partial \mathit{f(R)}}{\mathit{R}} > 0$ and $\frac{\partial^2 \mathit{f(R)}}{\mathit{R^2}} > 0$. The motivation for the assumption of these two conditions was the fact that  the authors' focus was to obtain solutions of relativistic extended objects with external matter fields and not vacuum black-hole solutions. However, the black-hole
solutions we obtain do not satisfy these conditions.

 The non-existence of the Birkhoff theorem in $\mathit{f(R)}$ gravitational theories are studied in \cite{PhysRevLett.53.315}. Recently, several authors have tried to investigate if the Birkhoff theorem is valid or not in the conformal frame \cite{Sotiriou:2011dz,Sebastiani:2010kv,PerezBergliaffa:2011gj,Gao:2016rdu,Amirabi:2015aya,Calza:2018ohl,Oliva:2011xu,Capozziello:2011wg}. In this study, we did not assume any approximation or carried out a conformal transformation to obtain the analytic solutions (\ref{ass1}). Our results confirm that the Birkhoff theorem is not valid for $\mathit{f(R)}$ gravity theories \cite{Xavier:2020ulw}. It is well known that the Birkhoff theorem is valid in GR due to the absence of spin-0 modes in the linearized field equations. When spin-0 is absent, the spherically symmetric spacetime cannot couple to higher-spin excitations \cite{Misner:1974qy,PhysRevLett.53.315}. Therefore, in the case of $\mathit{f(R)}$ gravitational theories, the differential equation satisfied by the Ricci scalar, $\mathit{R}$, plays the role of spin-0 modes. Hence, a non-trivial dependence between the metric and the Ricci scalar, in general, leads to the breaking of the Birkhoff theorem in $\mathit{f(R)}$. This is exactly the case of our analytic solution given by Eq. (\ref{R1}) which gives a non-trivial value of the Ricci scalar $\mathit{R}$.

We study the physics of those BHs by calculating the invariants of them and show that all the invariants behave up to the leading order as $\mathit{(R_{\mu \nu \rho \sigma} R^{\mu \nu \rho \sigma},R_{\mu \nu} R^{\mu \nu},R)}\approx\mathit{(\frac{1}{r^\mathrm{2}},\frac{1}{r^\mathrm{2}},\frac{1}{r^\mathrm{2}})}$ which are  different from  the Schwarzschild  BH which gives the leading term of the Kretschmann scalar as  $\frac{1}{r^6}$ and the other invariants  $\mathit{R_{\mu \nu} R^{\mu \nu}=R=\mathrm{const.}}$ This means  that the singularity of our BH for the Kretschmann scalar is much softer than that of GR.  We must emphasize that such merit is due to the contribution of the higher-order curvature of $f(R)$.

 To continue our investigation of these BHs we calculate the thermodynamical quantities like the Hawking temperature, entropy, quasi-local energy and Gibbs free energy. We show in detail the BH without the cosmological constant that all thermodynamical quantities are consistent with the literature. Essentially we show that the Hawking temperature depends on the degenerate horizon and if the temperature becomes less than the degenerate horizon we got a negative temperature and if it is greater we got a positive value. We repeat our calculations with the line-element that contains the cosmological constant and carried out our calculation for the AdS and dS spacetimes separately. We show that the degenerate horizons of those spacetimes play an important role to make the Hawking temperature has a positive value. Meanwhile, we show that our black satisfies the first law of thermodynamics.

Finally, we have studied the stability of these BHs. For this aim, we write the Lagrangian of the $\mathit{f(R)}$ gravitational theory as a scalar field that is coupled with the Ricci scalar.  Using the odd-type procedure we have derived the gradient instability condition and the radial propagation speed that is equal one for our BHs. Moreover, we have examined the stability conditions for those types of BH as shown in (\ref{stc}).

Finally, we would like to stress on the fact that our BH solutions given by Eq. (\ref{ass1n}) is not a general solution for the $\mathit{f(R)}$ gravitational  theory. This is because of the fact that in this study we have assumed $F=1+\frac{c_1}{r^2}$ to get the BH (\ref{ass1}). When we change the form of $F$ given by Eq. (\ref{ass1n}) we will get a new BH different from the one given by Eq. (\ref{ass1}). 
\begin{acknowledgments}
This work is partially supported  by MEXT KAKENHI Grant-in-Aid for Scientific Research on Innovative Areas ``Cosmic Acceleration'' No. 15H05890 (S.N.) and the JSPS Grant-in-Aid for Scientific Research (C) No. 18K03615 (S.N.).
\end{acknowledgments}
%

\end{document}